\documentclass[article]{jss}
\usepackage{amsthm, amsmath}

\DeclareMathOperator*{\argmax}{arg\,max}
\author{Hyungsuk Tak\\Harvard University \And 
             Joseph Kelly\\Google Inc.\And
             Carl Morris\\ Harvard University}
\title{\pkg{Rgbp}: An \proglang{R} Package for  Gaussian, Poisson, \\and Binomial  Random Effects Models \\with Frequency Coverage Evaluations}

\Plainauthor{Hyungsuk Tak, Joseph Kelly, Carl Morris} 
\Plaintitle{\pkg{Rgbp}: An R Package for  Gaussian, Poisson, and Binomial  Random Effects Modeling and Frequency Method Checking on Overdispersed Data} 
\Shorttitle{\pkg{Rgbp}: Random Effects Models, with Frequency Coverage Evaluations} 
\Abstract{
\pkg{Rgbp} is an \proglang{R} package that provides estimates and verifiable confidence intervals for random effects in two-level conjugate hierarchical models for overdispersed Gaussian, Poisson, and Binomial data.   \pkg{Rgbp} models aggregate data from $k$ independent groups summarized by observed sufficient statistics for each random effect, such as sample means, possibly with covariates. \pkg{Rgbp} uses approximate Bayesian machinery with unique improper priors for the hyper-parameters, which leads to good repeated sampling coverage properties for random effects. A special feature of  \pkg{Rgbp} is an option that generates  synthetic data sets to check whether the interval estimates for random effects actually meet the nominal confidence levels. Additionally, \pkg{Rgbp} provides inference statistics for the hyper-parameters, e.g., regression coefficients.}
\Plainkeywords{multilevel model, conjugate hierarchical generalized linear models, frequency method checking, coverage probability, shrinkage, R} 

\Address{
  Hyungsuk Tak\\
  Department of Statistics\\
  Harvard University\\
  1 Oxford Street, Cambridge, MA\\
  E-mail: \email{hyungsuk.tak@gmail.com}\\

  Joseph Kelly\\
  Google, Inc.\\
  76 Ninth Avenue, New York, NY\\
  E-mail: \email{josephkelly@google.com}\\

  Carl Morris\\
  Department of Statistics\\
  Harvard University\\
  1 Oxford Street, Cambridge, MA\\
  E-mail: \email{morris@stat.harvard.edu}\\
}

\begin{document}

\section[Introduction]{Introduction}
Gaussian, Poisson, or Binomial data from several independent groups sometimes have more variation than the assumed Gaussian, Poisson, or Binomial distributions of the first-level observed data. To account for this extra variability, called overdispersion, a  two-level conjugate hierarchical model regards first-level mean parameters as random effects that come from a population-level conjugate prior distribution. The main  goal of our two-level conjugate modeling is to estimate these random effects for a comparison between groups. For example, this model can be used to estimate unknown true batting averages (random effects) of baseball players for a comparison among players based on their numbers of hits and at-bats possibly with their covariate information. 


With an assumption of homogeneity within each group, the observed data are group-level aggregate data from $k$ independent  groups, composed of sufficient statistics for their $k$ random effects (without the population-level data). Specifically, the data for the Gaussian model consist of each group's sample mean and its standard error, those for the Poisson model use each group's outcome count and an exposure measure, and those for the Binomial model use the number of each group's  successful outcomes together with the total number of trials. For example, the observed data for the Binomial model can be the number of hits out of at-bats for each of $k$ baseball players, a sufficient statistic for the unknown true batting average of each player. The data analyzed by \pkg{Rgbp} may incorporate each group's covariate information, \emph{e.g.}, each player's position. 

These types of  data are common in various fields for estimating random effects. For example, biologists may be interested in unknown true tumor incidence rates in analyzing litter data composed of each litter's number of tumor-bearing animals out of  total number of animals at risk \citep{tamura1987stabilized}. The unknown true mortality rates on myocardial infarction can be estimated based on the death rate data collected from several independent clinical studies via a meta analysis \citep{gelman2014bayesian}.  County-level or state-level summary data in small area estimation problems \citep{ghosh1994small, rao2003small} can be used to estimate  population parameters, such as unknown unemployment rates.  

For such data, assuming homogeneity within each group, \pkg{Rgbp}'s two-level model may be viewed as a conjugate hierarchical generalized linear model \citep[HGLM,][]{lee1996hierarchical} where each random effect comes from a conjugate prior distribution.    However, the HGLM focuses on estimating regression coefficients to explore associations between covariates and observed data.   While \pkg{Rgbp} does that too, its emphasis concerns making valid point and interval estimates for the $k$ random effects for a comparison among groups.

A defining feature of \pkg{Rgbp} is to evaluate the repeated sampling coverage properties of the interval estimates for random effects \citep{morris1997, daniels1999prior, tang2002fitting, tang2011, morris2012}. This procedure distinguishes \pkg{Rgbp}  from  other \proglang{R} packages for similar hierarchical models, such as \pkg{hglm} \citep{ronnegaard2011hglm} for fitting conjugate hierarchical generalized models, because most software produces only estimation results without providing a quick way to evaluate their estimation procedures.  The evaluation procedure which we call \emph{frequency method checking} uses a parametric bootstrapping  method that generates synthetic data sets given the fitted values of the estimated hyper-parameters. The frequency method checking estimates the coverage rates of  interval estimates based on the simulated data sets and checks whether the estimated coverage rates achieve (or exceed) the nominal confidence level.

\pkg{Rgbp} combines Bayesian modeling tools with our  improper hyper-prior distributions  on the second-level parameters.  These hyper-prior distributions are known to produce good repeated sampling coverage rates for the Bayesian interval estimates for the $k$ random effects in two-level Gaussian models \citep{tang2011, morris2012, kelly2014advances}.  We extend these hyper-prior distributions for \pkg{Rgbp}'s Poisson and Binomial hierarchical models. 

For fitting the hierarchical model, \pkg{Rgbp} uses \emph{adjustment for density maximization} \citep[ADM, ][]{carl1988, morris1997, tang2011}.   ADM approximates a posterior density or a likelihood function by fitting a selected (one dimensional) Pearson family, based on the first two derivatives of the given density function.  For example, when the Normal distribution is the chosen Pearson family, ADM reduces to a Delta method via maximum likelihood estimation.  Besides ADM, \pkg{Rgbp} provides an option for the Binomial  hierarchical model to draw independent posterior samples of random effects and hyper-parameters via an acceptance-rejection method \citep{robert2013monte}.

The rest of this paper is organized as follows. We specify the Bayesian hierarchical models and discuss their posterior propriety in Section \ref{sec2}. In Section \ref{inference}, we explain the inferential models used to estimate the model parameters. We describe the estimation procedures including  ADM and the acceptance-rejection method in Section  \ref{sec3} and \ref{sec_accept}, respectively. We introduce  frequency method checking techniques in Section \ref{sec4}.  We explain the basic usages of \pkg{Rgbp}'s main functions with three examples in Section \ref{sec6}, and specify the functions further with various options in Section \ref{sec5}.

\section[Hierarchical Structure]{Conjugate hierarchical modeling structure} \label{sec2}

\pkg{Rgbp} allows users to choose one of three hierarchical models according to the type of data, namely Normal-Normal, Poisson-Gamma, and Binomial-Beta models. Although there are more hierarchical models, we choose the three models because these are based on the most common types of  data we may encounter in practice. Also, their conjugacy leads to linear posterior means simplifying computations.

Our parametrization of the three hierarchical models leads to an intuitive shrinkage interpretation in inference because the shrinkage factors under our parametrization are determined by the relative amount of information in the prior compared to the data \citep{morris1983natural}. 
 
\subsection[Normal-Normal]{Normal-Normal model for Gaussian data}
The following Normal-Normal hierarchical model (hereafter the Gaussian model) assumed by \pkg{Rgbp} is useful when the group-level aggregate data from $k$ independent  groups are continuous (or approximately continuous) variables with known standard errors. The subscript \emph{j} below indicates the \emph{j}th group among \emph{k} groups in the dataset. For $j=1, 2, \ldots, k$, 
\begin{align}
y_{j}\mid \mu_{j} & \stackrel{\textrm{indep.}}{\sim} \textrm{Normal}\!\left(\mu_{j}, V_{j}\right)\!,\label{normalobs}\\
\mu_{j}\mid \boldsymbol{\beta}, A & \stackrel{\textrm{indep.}}{\sim} \textrm{Normal}\!\left(\mu^E_{j}, A\right)\!,\label{normalprior}
\end{align}

where $y_j$ is an observed unbiased estimate, \emph{e.g.}, sample mean, for random effect $\mu_j$, $V_{j}$ is a completely known standard error of $y_j$, $\mu^E_{j}$ is an expected random effect defined as ${\E(\mu_j\mid \boldsymbol{\beta}, A)} = \boldsymbol{x}_j^\top \boldsymbol{\beta}=\beta_{1}x_{j, 1}+\beta_{2}x_{j, 2} + \cdots + \beta_{m}x_{j, m}$, and $m$ is the number of unknown regression coefficients.   It is assumed that the second-level variance $A$ is unknown and that the vector of $m$  regression coefficients $\boldsymbol{\beta}$ is also unknown unless otherwise specified. If no covariates are available, but with an unknown intercept term, then $\boldsymbol{x}_j^\top \boldsymbol{\beta}=\beta_1$ ($m=1$) and thus $\mu^E_{j}=\mu^E=\beta_1$ for all $j$, resulting in an exchangeable conjugate prior distribution for the random effects.  Based on these conjugate prior distributions for random effects, it is easy to derive the conditional posterior distribution of each random effect. For $j=1, 2, \ldots, k$,

\begin{equation} \label{normalpost}
\mu_{j}\mid  \boldsymbol{\beta}, A, \boldsymbol{y} \stackrel{\textrm{indep.}}{\sim}\textrm{Normal}\!\left((1-B_{j})y_{j} + B_{j}\mu^E_{j},~(1-B_{j})V_{j}\right)\!,
\end{equation}
where $B_{j}\equiv V_{j}/(V_{j} + A)$ is a shrinkage factor of group $j$ and $\boldsymbol{y}=(y_1, y_2, \ldots, y_k)^\top$. The conditional posterior mean $\E(\mu_{j}\mid \boldsymbol{\beta}, A, \boldsymbol{y} )$, denoted by $\mu^\ast_j$,  is a convex combination of the observed sample mean $y_j$ and the expected random effect $\mu^E_j$ weighted by the shrinkage factor $B_j$. If the variance of the conjugate prior distribution, $A$, is smaller than the variance of the observed distribution, $V_j$, then we expect the posterior mean to borrow more information from the second-level conjugate prior distribution.

\subsection[Poisson-Gamma]{Poisson-Gamma model for Poisson data}\label{poissonsubsec}
\pkg{Rgbp} can estimate a conjugate Poisson-Gamma hierarchical model (hereafter the Poisson model) when the group-level aggregate data from $k$ independent  groups consist of non-negative count data without upper limit. However, its usage is limited to the case where the expected random effect, $\lambda^E_j=\exp(\boldsymbol{x}_j^\top \boldsymbol{\beta})$,  is known  (or equivalently all the regression coefficients $\boldsymbol{\beta}$ are known ($m=0$)); we may be able to obtain this information from the past studies or from experts. For $j=1, 2, \ldots, k$, 
\begin{align}
y_j\mid \lambda_j &\stackrel{\textrm{indep.}}{\sim}  \textrm{Poisson}\!\left(n_{j}\lambda_{j}\right)\!,\\
\lambda_{j}\mid r &\stackrel{\textrm{indep.}}{\sim} \textrm{Gamma}\!\left(r\lambda^E_j, r\right)\!,\label{gamma_prior}
\end{align}
where $y_j$ is the number of events happening, $n_{j}$ is the exposure of group $j$, which is not necessarily an integer, $\lambda^E_j=\E(\lambda_j\mid r)$ is the known expected random effect ($m=0$), and $r$ is the unknown second-level variance component. The mean and variance of this conjugate Gamma prior distribution are $\lambda^E$ and $\lambda^E/r$, respectively\footnote{The density function of this Gamma prior distribution in \eqref{gamma_prior} is $f(\lambda_j\mid r)\propto \lambda_j^{r\lambda^E_j - 1}\exp(-r\lambda_j)$}.     \cite{albert1988computational} interprets $r$ as the amount of prior information as $n_{j}$ represents the amount of observed information because the uncertainty of the conjugate prior distribution increases as $r$ decreases and vice versa. The conditional posterior distribution of the random effect $\lambda_j$ for this Poisson model is
\begin{equation} \label{gamma_post}
\lambda_j\mid r, \boldsymbol{y} \stackrel{\textrm{indep.}}{\sim}\textrm{Gamma}\!\left(r\lambda^E_j + n_j\bar{y}_{j},~ r + n_j\right)\!,
\end{equation}  
where $\bar{y}_j\equiv y_j/n_j$. The mean and variance of the conditional posterior distribution are
\begin{equation}\label{gammapost_mean_var}
\lambda^\ast_j\equiv \E(\lambda_j\mid r, \boldsymbol{y} )=(1-B_{j})\bar{y}_{j} + B_{j}\lambda^E_j~~\textrm{and}~~\VAR(\lambda_j\mid r, \boldsymbol{y})=\frac{\lambda^\ast_j}{r+n_j}.
\end{equation}
where $B_{j}\equiv r / (r+n_{j})$ is the shrinkage factor of group $j$, the relative amount of information in the prior compared to the data. The conditional posterior mean is  a convex combination of $\bar{y}_{j}$ and  $\lambda^E_j$ weighted by $B_j$. If the conjugate prior distribution contains more information than the observed data have, \emph{i.e.}, ensemble sample size $r$ exceeds individual sample size $n_{j}$, then the posterior mean shrinks  towards the prior mean by more than 50\%.

The conditional posterior variance in  \eqref{gammapost_mean_var} is linear in the conditional posterior mean, whereas a slightly different parameterization for a Poisson-Gamma model has been used elsewhere \citep{morris1997} that makes the variances  quadratic functions of means. 

\subsection[Binomial-Beta]{Binomial-Beta model for Binomial data}
\pkg{Rgbp} can fit a conjugate Binomial-Beta hierarchical model (hereafter the Binomial model) when the group-level aggregate data from $k$ independent  groups are composed of each group's number of successes out of total number of trials. The expected random effect in the Binomial model is either known ($m=0$) or unknown ($m\ge1$). For $j=1, 2, \ldots, k$,
\begin{align}
y_{j} \mid p_{j} &\stackrel{\textrm{indep.}}{\sim} \textrm{Binomial}(n_{j}, ~p_{j}),\\
p_{j} \mid \boldsymbol{\beta}, r &\stackrel{\textrm{indep.}}{\sim} \textrm{Beta}\!\left(rp^E_j,~ r(1-p^E_j)\right)\!,
\end{align}

where $y_{j}$  is the number of successes  out of $n_{j}$ trials, $p^E_j$ is the expected random effect  of group $j$ defined as $p^E_j\equiv \E(p_j\mid \boldsymbol{\beta}, r)=\exp(\boldsymbol{x}_j^\top\boldsymbol{\beta})/(1+\exp(\boldsymbol{x}_j^\top\boldsymbol{\beta}))$. The  vector of the $m$ logistic regression coefficients $\boldsymbol{\beta}$ and the second-level variance component $r$ are unknown. The mean and variance of the conjugate Beta prior distribution for group $j$ are $p^E_j$ and $p^E_j(1-p^E_j)/(r+1)$, respectively.  The resultant conditional posterior distribution of random effect $p_j$  is
\begin{equation} \label{beta_post}
p_{j}\mid  \boldsymbol{\beta}, r, \boldsymbol{y} \stackrel{\textrm{indep.}}{\sim}\textrm{Beta}\!\left(n_{j}\bar{y}_{j}+rp^E_j,~n_{j}(1-\bar{y}_{j})+r(1-p^E_j)\right)\!,
\end{equation}
where $\bar{y}_j=y_j/n_j$ is the observed proportion of group $j$. The mean and variance of the conditional posterior distribution are
\begin{equation}\label{betapost_mean_var}
p_j^\ast\equiv \E(p_j\mid \boldsymbol{\beta}, r, \boldsymbol{y} )=(1-B_{j})\bar{y}_{j} + B_{j}p^E_j~~\textrm{and}~~\VAR(p_j\mid  \boldsymbol{\beta}, r,  \boldsymbol{y})=\frac{p_j^\ast(1-p_j^\ast)}{r+n_j+1}.
\end{equation}
The conditional posterior mean $p_j^\ast$ is  a convex combination of $\bar{y}_{j}$ and $p^E_j$ weighted by $B_j\equiv r / (r + n_j)$ like the Poisson model.  If the conjugate prior distribution contains more information than the observed distribution does ($r>n_j$), then the resulting conditional posterior mean borrows more information from the conjugate Beta prior distribution.

\subsection[Hyper-prior distributions]{Hyper-prior distribution}
Hyper-prior distributions are the distributions assigned to the second-level parameters called hyper-parameters. Our choices for the hyper-prior distributions are
\begin{equation}\label{eq:hyper}
\boldsymbol{\beta} \sim \textrm{Uniform on}~ \textbf{R}^{m}~~\textrm{and}~~A \sim \textrm{Uniform}(0, \infty) ~~\left(\textrm{or} ~1/r\sim \textrm{Uniform}(0, \infty)\right)\!.
\end{equation}
The improper flat hyper-prior distribution on $\boldsymbol{\beta}$ is a common non-informative choice.  In the Gaussian case, the flat hyper-prior distribution on the second-level variance $A$ is known to produce good repeated sampling coverage properties of the Bayesian interval estimates for the random effects \citep{tang2011, morris2012, kelly2014advances}. The resulting full posterior distribution of the random effects and hyper-parameters is proper if $k\ge m+3$ \citep{tang2011, kelly2014advances}. 

In the other two cases, Poisson and Binomial, the flat prior distribution on $1/r$ induces the same improper prior distribution on shrinkages ($d B_j/B_{j}^{2} $) as does $A$ with the Uniform($0, \infty$) for the Gaussian case. The Poisson model with this hyper-prior distribution on $r$, \emph{i.e.}, $dr/r^2$,  provides posterior propriety if there are at least two groups whose observed values $y_j$ are non-zero  and the expected random effects, $\lambda^E_j$, are known ($m=0$); see Appendix \ref{propriety_poisson} for its proof. If $\lambda^E_j$ is unknown, \pkg{Rgbp} cannot yet give reliable results because we have not verified posterior propriety.  If the Poisson is being used as an approximation to the Binomial and the exposures are known integer values, then we recommend using the Binomial model with the same hyper-prior distributions.

As for posterior propriety of the Binomial model, let us define an \emph{interior group} as the group whose number of successes $y_j$ are neither 0 nor $n_j$, and $k_y$ as the number of interior groups among $k$ groups. The full posterior distribution of random effects and hyper-parameters is proper if and only if there are at least two interior groups in the data and the $k_y\times m$ covariate matrix of the interior groups is of full rank $m$ ($k_y\ge m$) \citep{tak2016propriety}.

\section[Inference]{The inferential model}\label{inference}

The likelihood function of hyper-parameters, $A$ and $\boldsymbol{\beta}$, for the Gaussian  model is derived from the independent Normal distributions of the observed data with random effects integrated out. 
\begin{equation}\label{marginal_normal}
L(A, \boldsymbol{\boldsymbol{\beta}})=\prod_{j=1}^k f(y_j\mid A, \boldsymbol{\boldsymbol{\beta}})=\prod_{j=1}^k \frac{1}{\sqrt{2\pi (A+V_j)}}\exp\left(-\frac{(y_j-\mu^E_j)^2}{2(A+V_j)}\right).
\end{equation}
The joint posterior density  function of hyper-parameters for the Gaussian  model is proportional to their likelihood function in  \eqref{marginal_normal} because we use flat improper hyper-prior density functions for $A$ and $\boldsymbol{\beta}$:
\begin{equation}\label{marginal_post_normal}
f(A, \boldsymbol{\boldsymbol{\beta}}\mid \boldsymbol{y})\propto L(A, \boldsymbol{\boldsymbol{\beta}}).
\end{equation}

The likelihood function of $r$ for the Poisson  model  comes from the independent Negative-Binomial  distributions of the observed data with the random effects integrated out.
\begin{equation}\label{marginal_poisson}
L(r)=\prod_{j=1}^k f(y_j\mid r)=\prod^{k}_{j=1} \frac{\Gamma(r \lambda^E_j+y_j)}{\Gamma(r\lambda^E_j)(y_j!)}(1-B_{j})^{y_{i}}B_{j}^{r \lambda^E_j},
\end{equation}
where $\Gamma(a)$ is a gamma function defined as $\int_0^\infty x^{a-1}\exp(x)dx$ for a positive constant $a$. The posterior density function of $r$  for the Poisson  model is  the likelihood function in  \eqref{marginal_poisson} times the hyper-prior density function of $r$, \emph{i.e.}, $dr/r^2$:
\begin{equation}\label{marginal_post_poisson}
f(r \mid \boldsymbol{y})\propto L(r)/r^2.
\end{equation}

The likelihood function of hyper-parameters $r$ and $\boldsymbol{\beta}$ for the Binomial  model is derived from the independent Beta-Binomial  distributions of the observed data with random effects integrated out \citep{skellam1948}.
\begin{equation}\label{marginal_binomial}
L(r, \boldsymbol{\boldsymbol{\beta}})=\prod_{j=1}^k f(y_j\mid r, \boldsymbol{\boldsymbol{\beta}})=\prod_{j=1}^k\binom{n_j}{y_j}\frac{B(y_j+rp^E_j, ~n_j-y_j+r(1-p^E_j))}{B(rp^E_j, ~r(1-p^E_j))},
\end{equation}
where the notation $B(a, b)~(\equiv\int_0^1 v^{a-1}(1-v)^{b-1}dv)$ indicates a beta function for positive constants $a$ and $b$. The joint posterior density  function of hyper-parameters $f(r, \boldsymbol{\boldsymbol{\beta}}\mid \boldsymbol{y})$ for the Binomial  model is proportional to their likelihood function in \eqref{marginal_binomial} multiplied by the hyper-prior density functions of $r$ and $\boldsymbol{\beta}$ based on distributions in  \eqref{eq:hyper}:
\begin{equation}\label{marginal_post_binomial}
f(r, \boldsymbol{\boldsymbol{\beta}}\mid \boldsymbol{y})\propto L(r, \boldsymbol{\boldsymbol{\beta}})/r^2.
\end{equation}


Our goal is to obtain the point and interval estimates of the random effects from their joint unconditional posterior density which, for the Gaussian  model, can be expressed as 
\begin{equation}\label{mcintegration_normal}
f(\boldsymbol{\mu}\mid \boldsymbol{y})=\int  f(\boldsymbol{\mu}\mid A, \boldsymbol{\beta}, \boldsymbol{y}) f(A, \boldsymbol{\beta}\mid \boldsymbol{y})dA d\boldsymbol{\beta},
\end{equation}
where $\boldsymbol{\mu}=(\mu_1, \mu_2, \ldots, \mu_k)^\top$ and the distributions in the integrand  are given in \eqref{normalpost} and \eqref{marginal_post_normal}.  For the Poisson  model, the joint unconditional posterior density for the random effects is
\begin{equation}\label{mcintegration_poisson}
f(\boldsymbol{\lambda}\mid \boldsymbol{y})=\int  f(\boldsymbol{\lambda}\mid r, \boldsymbol{y}) f(r\mid \boldsymbol{y})dr,
\end{equation}
where  $\boldsymbol{\lambda}=(\lambda_1, \lambda_2, \ldots, \lambda_k)^\top$ and  the distributions in the integrand  are given in \eqref{gamma_post} and \eqref{marginal_post_poisson}. For the Binomial  model, the joint unconditional posterior density for the random effects is
\begin{equation}\label{mcintegration_binomial}
f(\boldsymbol{p}\mid \boldsymbol{y})=\int  f(\boldsymbol{p}\mid r, \boldsymbol{\beta}, \boldsymbol{y}) f(r, \boldsymbol{\beta}\mid \boldsymbol{y})dr d\boldsymbol{\beta},
\end{equation}
where $\boldsymbol{p}=(p_1, p_2, \ldots, p_k)^\top$ and  the distributions in the integrand  are given in \eqref{beta_post} and \eqref{marginal_post_binomial}.

\section[Estimation]{Estimation via the adjustment for density maximization}\label{sec3}

We illustrate our estimation procedure which utilizes adjustment for density maximization  \citep[ADM,][]{carl1988, morris1997, tang2011}. ADM is a method  to  approximate a distribution by a member of Pearson family of distributions and obtain moment estimates via maximization. The ADM procedure for the Gaussian model adopted in \pkg{Rgbp} is well documented in \cite{kelly2014advances} and thus  we  describe  the ADM procedure  using the Poisson and Binomial model in this section. 

\subsection[shrinkage]{Estimation for shrinkage factors and expected random effects}\label{shrinkage}

Our goal here is to estimate the unconditional posterior moments of the shrinkage factors and the expected random effects because they are used to estimate the unconditional posterior moments of the random effects. 

\subsubsection{Unconditional posterior moments of shrinkage factors}  It is noted that  the shrinkage factors  are a function of $r$, \emph{i.e.}, $B_{j}= B_{j}(r)=r/(r+n_{j})$ (or a function of $A$ for the Gaussian model). A common method of estimation of $B_{j}$ is to approximate the likelihood of $r$ with two derivatives and use  a Delta method for an asymptotic Normal distribution of $\hat{B}_{j}(\hat{r}_{MLE})$. This Normal approximation, however, is defined on $(-\infty, \infty)$ whereas $B_{j}$ lies on the unit interval between 0 and 1, and hence in small sample sizes the Delta method can result in point estimates lying on the boundary of the parameter space, from which the restricted MLE procedure sometimes suffers \citep{tang2011}.

To continue with a maximization-based estimation procedure but to steer clear of aforementioned boundary issues we make use of  ADM. The ADM approximates the distribution of the function of the parameter of interest by one of the Pearson family distributions using the first two derivatives as the Delta method does; the Delta method is a special case of the ADM based on the Normal distribution. 


The ADM procedure specified in \cite{tang2011} assumes that the unconditional posterior distribution of a shrinkage factor follows a Beta distribution; for $j=1, 2, \ldots, k$,
\begin{equation}\label{admshrinkage}
B_j\mid\boldsymbol{y}\sim \textrm{Beta}(a_{1j},~ a_{0j}).
\end{equation}

The mean of Beta distribution $a_{1j}/(a_{1j}+a_{0j})$ is not the same as its mode $(a_{j1}-1)/(a_{j1}+a_{j0}-2)$. The ADM works on an adjusted posterior distribution $f^A(B_j\mid \boldsymbol{y})\propto B_j(1-B_j)f(B_j\vert \boldsymbol{y})$ so that the mode of $f^A(B_j\mid \boldsymbol{y})$ is the same as the mean of the original Beta distribution. The assumed posterior mean and variance of the $j$th shrinkage factor are
\begin{align}
\E(B_j\mid\boldsymbol{y})&=\frac{a_{1j}}{a_{1j}+a_{0j}}=\argmax_{B_j}~ f^A(B_j\mid \boldsymbol{y})\equiv B'_j,\label{admmean}\\
\VAR(B_j\mid\boldsymbol{y})&=\frac{B'_j(1-B'_j)}{a_{1j}+a_{0j}+1}=\frac{B'_j(1-B'_j)}{B'_j(1-B'_j)\left[-\frac{d^2}{dB^2_j}\log\left(f^A(B_j\mid \boldsymbol{y})\right)~\bigg\vert_{B_j=B'_j}\right] +1}.\label{admvar}
\end{align}

The ADM estimates these mean and variance using the marginal posterior distribution of $r$, $f(r\mid\boldsymbol{y})\propto L(r)/r^2$. The marginal likelihood, $L(r)=\int L(\boldsymbol{\beta}, r)d\boldsymbol{\beta}$, for the Binomial model is obtained via the Laplace approximation with a Lebesque measure on $\boldsymbol{\beta}$ and that for the Poisson model is specified in  \eqref{marginal_poisson}. 

Considering that  \eqref{admmean} involves maximization and  \eqref{admvar} involves  calculating the second derivative of $\log(f^A(B_j\mid \boldsymbol{y}))$, we work on a logarithmic scale of $r$, \emph{i.e.}, $\alpha=-\log(r)$ (or $\alpha=\log(A)$ for the Gaussian model). This is because the distribution of $\alpha$ is more symmetric than that of $r$ and $\alpha$ is defined on a real line without any boundary issues. Because $f^A(B_j\mid\boldsymbol{y})$ is proportional to the marginal posterior density $f(\alpha\mid\boldsymbol{y})\propto \exp(\alpha) L(\alpha)$ as shown in \cite{tang2011}, the  posterior mean in  \eqref{admmean} is estimated by
\begin{equation}\label{meaninvariance}
\hat{B}'_j=\frac{\exp(-\hat{\alpha})}{n_j+\exp(-\hat{\alpha})},
\end{equation}


where $\hat{\alpha}$ is the mode of $f(\alpha\mid\boldsymbol{y})$, \emph{i.e.}, $\argmax_{\alpha}\{\alpha+\log(L(\alpha))\}$ .

To estimate the variance in \eqref{admvar}, \cite{tang2011} introduced the invariant information defined as \begin{eqnarray}
I_\textrm{inv} &\equiv& -\frac{d^2 \log(f^A(B_j\mid \boldsymbol{y}))}{d[\textrm{logit}(B_j)]^2}~\bigg\vert_{B_j=\hat{B}'_j}=-\frac{d^2 \log(f^A(B_j(r)\mid \boldsymbol{y}))}{d[\log(r)]^2}~\bigg\vert_{r=\hat{r}}\label{invariance}\\
&=& -\frac{d^2 \log(f^A(B_j(r(\alpha))\mid \boldsymbol{y}))}{d\alpha^2}~\bigg\vert_{\alpha=\hat{\alpha}}.\nonumber
\end{eqnarray}

This invariant information is  the negative second-derivative of $\alpha+\log(L(\alpha))$ evaluated at the mode $\hat{\alpha}$. Using the invariant information, we  estimate the unconditional posterior variance of shrinkage factor in  \eqref{admvar} by
\begin{equation}\label{varianceinvariance}
\widehat{\VAR}(B_j\mid\boldsymbol{y})=\frac{(\hat{B}'_j)^2(1-\hat{B}'_j)^2}{I_\textrm{inv} +\hat{B}'_j(1-\hat{B}'_j)}.
\end{equation}

We obtain the estimates of $a_{1j}$ and $a_{0j}$, the two parameters of the Beta distribution in  \eqref{admshrinkage}, by matching them to the estimated unconditional posterior mean and variance of $B_j$ specified in  \eqref{meaninvariance} and \eqref{varianceinvariance} as follows.
\begin{equation}\label{admpara}
\hat{a}_{1j}=\frac{I_\textrm{inv}}{1-\hat{B}'_j}~~\textrm{and}~~\hat{a}_{0j}=\frac{I_\textrm{inv}}{\hat{B}'_j}.
\end{equation}

The moments of the Beta distribution are well defined as a function of $a_{1j}$ and $a_{0j}$, \emph{i.e.},  $\E(B^c_j\mid\boldsymbol{y})=B(a_{1j} + c, a_{0j})/B(a_{1j}, a_{0j})$ for $c\ge0$. Their estimates are 
\begin{equation}\label{shrinkagemoments}
\widehat{\E}(B^c_j\mid\boldsymbol{y})= \frac{B(\hat{a}_{1j} + c,~ \hat{a}_{0j})}{B(\hat{a}_{1j},~ \hat{a}_{0j})}.
\end{equation}

The ADM approximation to the shrinkage factors via Beta distributions is empirically proven to be more accurate than a Laplace approximation \citep{carl1988, morris1997, tang2011, morris2012}.

\subsubsection{Unconditional posterior moments of expected random effects}
We estimate the unconditional posterior moments of expected random effects using their relationship to the conditional posterior moments. For a non-negative constant $c$, the unconditional posterior moments are
\begin{equation}
\E\!\left(\left(p^E_j\right)^c\big\vert~\boldsymbol{y}\right)=\E\!\left[\E\!\left(\left(p^E_j\right)^c\big\vert~\alpha, \boldsymbol{y}\right)\big\vert~\boldsymbol{y}\right]\!.
\end{equation}
We approximate the unconditional posterior moments on the left hand side by the conditional posterior moments with $\hat{\alpha}$ inserted \citep{kass1989approximate}, \emph{i.e.}, by $\E((p^E_j)^c\mid \hat{\alpha}, \boldsymbol{y})$.

However, calculating conditional posterior moments of each expected random effect involves an intractable integration. For example, the first conditional posterior moment of $p^E_j$ is
\begin{equation}\label{priormoment1}
\E(p^E_{j}\mid\hat{\alpha}, \boldsymbol{y})=\E\bigg(\frac{\exp(x_j^\top\boldsymbol{\beta})}{1+\exp(x_j^\top\boldsymbol{\beta})}\bigg\vert\hat{\alpha}, \boldsymbol{y}\bigg)=\int_{\mathbf{R}^{m}} \frac{\exp(x_j^\top\boldsymbol{\beta})}{1+\exp(x_j^\top\boldsymbol{\beta})}f(\beta\mid\hat{\alpha}, \boldsymbol{y})d\boldsymbol{\beta}.
\end{equation}
Thus, we use another ADM, assuming the conditional posterior distribution of each expected random effect is a Beta distribution as follows.
\begin{equation}\label{admpriormean}
p^E_{j}\mid\hat{\alpha}, \boldsymbol{y}=\frac{\exp(x_j^\top\boldsymbol{\beta})}{1+\exp(x_j^\top\boldsymbol{\beta})}\bigg\vert\hat{\alpha}, \boldsymbol{y}\sim \textrm{Beta}(b_{1j},~ b_{0j})\sim \frac{G_1}{G_1+G_0},
\end{equation}
where $G_1$ is a random variable following a Gamma$(b_{1j}, 1)$ distribution and independently $G_0$ has a Gamma($b_{0j}, 1)$ distribution.  The representation in  \eqref{admpriormean} is equivalent to  $\exp(x_j^\top\boldsymbol{\beta})\vert\hat{\alpha}, \boldsymbol{y}\sim G_1/G_0$, a ratio of two independent Gamma random variables. Its mean and variance are
\begin{align}
\E(\exp(x_j^\top\boldsymbol{\beta})\mid\hat{\alpha}, \boldsymbol{y})&=\E\bigg(\frac{G_1}{G_0}\bigg)=\frac{b_{1j}}{b_{0j}-1}\equiv\eta_j,\label{priormean1}\\
~~~~~\VAR(\exp(x_j^\top\boldsymbol{\beta})\mid\hat{\alpha}, \boldsymbol{y})&=\VAR\bigg(\frac{G_1}{G_0}\bigg)=\frac{\eta_j(1+\eta_j)}{b_{0j}-2}.\label{priorvariance1}
\end{align}

To estimate $b_{1j}$ and $b_{0j}$, we assume that the conditional posterior distribution of $\boldsymbol{\beta}$ given $\hat{\alpha}$ and $\boldsymbol{y}$ follows a Normal distribution with mean $\hat{\boldsymbol{\beta}}$ and variance-covariance matrix $\hat{\Sigma}$, where $\hat{\boldsymbol{\beta}}$ is the mode of $f(\boldsymbol{\beta}\mid \hat{\alpha}, \boldsymbol{y})$  and $\hat{\Sigma}$ is an inverse of the negative Hessian matrix at the mode. Thus, the posterior distribution of $x_j^\top\boldsymbol{\beta}$ is also Normal with mean $x_j^\top\hat{\boldsymbol{\beta}}$ and variance $x_j^\top\hat{\Sigma} x_j$.

Using the property of the log-Normal distribution for $\exp(x_j^\top\boldsymbol{\beta})$, we estimate the posterior mean and variance in  \eqref{priormean1} and \eqref{priorvariance1} as
\begin{align}
\hat{\E}(\exp(x_j^\top\boldsymbol{\beta})\mid\hat{\alpha}, \boldsymbol{y})&=\exp\!\left(x_j^\top\hat{\boldsymbol{\beta}}+x_j^\top\hat{\Sigma} x_j/2\right)=\hat{\eta}_j,\label{priormean2}\\
\widehat{\VAR}(\exp(x_j^\top\boldsymbol{\beta})\mid\boldsymbol{y})&=\hat{\eta}^2_j\!\left(\exp(x_j^{T}\hat{\Sigma} x_j)-1\right)\!.\label{priorvariance2}
\end{align}

We estimate the values of $b_{1j}$ and $b_{0j}$  by matching them to the estimated unconditional posterior mean and variance of $\exp(x_j^\top\boldsymbol{\beta})$ in  \eqref{priormean2} and \eqref{priorvariance2}, that is,
\begin{equation}\label{priormeanpara}
\hat{b}_{1j}=\hat{\eta}_j(\hat{b}_{0j}-1)~~\textrm{and}~~\hat{b}_{0j}=\frac{1+\hat{\eta}_j}{\hat{\eta}_j\left(\exp(x_j^{T}\hat{\Sigma} x_j)-1\right)}+2.
\end{equation}

Finally, we estimate the unconditional posterior moments of the expected random effects by
\begin{equation}\label{est_exp_randomeffect}
 \hat{\E}\left(\left(p^E_{j}\right)^c\mid\hat{\alpha}, \boldsymbol{y}\right)=\frac{B\left(\hat{b}_{1j} + c, \hat{b}_{0j}\right)}{B\left(\hat{b}_{1j}, \hat{b}_{0j}\right)} \textrm{ for } c\ge0.
\end{equation}

The ADM approximation to a log-Normal density via a \emph{F}-distribution (represented by a ratio of two independent Gamma random variables) is known to be more accurate than the Laplace approximation \citep{carl1988}.


For the Gaussian model \citep{tang2011}, the conditional posterior distribution of $\boldsymbol{\beta}$ given $\hat{A}$ and $\boldsymbol{y}$ is Normal whose mean and variance-covariance matrix are
\begin{equation}
\left(X^\top D^{-1}_{V+\hat{A}} X\right)^{-1}X^\top D^{-1}_{V+\hat{A}}\boldsymbol{y}~~ \textrm{and}~~ \left(X^\top D^{-1}_{V+\hat{A}} X\right)^{-1},
\end{equation}
respectively, where $X\equiv (\boldsymbol{x}_1, \boldsymbol{x}_2, \ldots, \boldsymbol{x}_k)^\top$ is a $k \times m$ covariate matrix and $D_{V+\hat{A}}$ is a $k \times k$ diagonal matrix with the $j$-th diagonal element equal to $V_j+\hat{A}$. Because $\boldsymbol{x}^\top\boldsymbol{\beta}$ given $\hat{A}$ and $\boldsymbol{y}$ is also Normally distributed, we easily obtain the conditional posterior moments of $\mu^E_j=\boldsymbol{x}^\top\boldsymbol{\beta}$ given $\hat{A}$ and use them to estimate unconditional posterior moments of $\mu^E_j$.

\subsection{Estimation for random effects}
We illustrate how we obtain approximate unconditional posterior distributions of random effects using the estimated unconditional posterior moments of shrinkage factors and those of expected random effects.
 It is intractable to derive analytically the unconditional posterior
distribution of each random effect for the three models. Thus, we approximate the
distributions by matching the estimated posterior moments with a skewed-Normal distribution \citep{azzalini1985class} for
the Gaussian model, a Gamma distribution for the Poisson model, and a Beta distribution for the Binomial model; for $j=1, 2, \ldots, k$,
\begin{align}
\mu_j\mid\boldsymbol{y} &\sim \textrm{skewed-Normal}(\phi,~ \omega,~ \delta),\label{skewnormal}\\
\lambda_j\mid \boldsymbol{y} &\sim \textrm{Gamma}(s_{1j},~ s_{0j}),\label{gammapost}\\
p_j\mid \boldsymbol{y} &\sim \textrm{Beta}(t_{1j},~ t_{0j}),\label{betapost}
\end{align}
where $(\phi, \omega, \delta)$ of the skewed-Normal distribution  are location,  scale, and skewness parameters, respectively.

\cite{morris2012} first noted that the unconditional posterior distribution of
the random effect in a two-level conjugate Gaussian model might be
skewed. \cite{kelly2014advances}  shows that the skewed-Normal approximation to
the unconditional posterior distribution of the random effect is better than a
Normal approximation ($\mu_j\mid\boldsymbol{y}\sim\textrm{Normal}$) in terms
of the repeated sampling coverage properties of random
effects. \cite{kelly2014advances} estimates the first three moments of
the random effects by noting that $\mu_j$ is Normally
distributed given $A$ and  $\boldsymbol{y}$, and thus estimates the moments by using the ADM approximation of
the shrinkage factors, $B_{j}$, and the law of third cumulants \citep{brill}. The
three estimated moments are then matched to the first three moments of the
skewed-Normal distribution, \emph{i.e.}, $\E(\mu_j\mid \boldsymbol{y})
=\phi+\omega\delta\sqrt{2/\pi}$, $\VAR(\mu_j\mid
\boldsymbol{y})=\omega^2(1-2\delta^2/\pi)$, and Skewness$(\mu_j\mid
\boldsymbol{y})=(4-\pi)\delta^3/[2(\pi/2-\delta^2)^{3/2}]$
\citep{azzalini1985class}. The full derivation can be found in \cite{kelly2014advances}.

The unconditional posterior mean and variance of random effect $\lambda_j$ in the Poisson model are
\begin{align}
\E(\lambda_j\mid \boldsymbol{y}) &=\E(\E(\lambda_{j}\mid r, \boldsymbol{y})\mid\boldsymbol{y})=(1-\E(B_j\mid\boldsymbol{y}))\bar{y}_j + \E(B_j\mid  \boldsymbol{y})\lambda^E_j,\label{postmean_poisson} \\
\VAR(\lambda_{j}\mid \boldsymbol{y}) &=  \E(\VAR(\lambda_{j}\mid r,  \boldsymbol{y})\mid \boldsymbol{y})+\VAR(\E(\lambda_{j}\mid r, \boldsymbol{y})\mid \boldsymbol{y})\label{postvar_poisson}\\
&= \E\big(\lambda^{\ast}_{j}/ (r+n_{j})\mid  \boldsymbol{y}\big)+\VAR\big(B_{j}(\bar{y}_{j}-\lambda^E_{j})\mid \boldsymbol{y}\big) \\
&=\left[\bar{y}_j\E\left((1-B_j)^2\vert \boldsymbol{y}\right)+\lambda^E_j\E\left((1-B_j)B_j\vert \boldsymbol{y}\right)\right]/n_j+(\bar{y}_{j}-\lambda^E_{j})^2\VAR\big(B_{j}\vert \boldsymbol{y}\big).\label{approximate_var_poisson}
\end{align}
To estimate these, we insert  the estimated unconditional posterior moments of shrinkage factors in \eqref{shrinkagemoments} into both \eqref{postmean_poisson} and \eqref{approximate_var_poisson}. Let  $\hat{\mu}_{\lambda_j}$ and $\hat{\sigma}^2_{\lambda_j}$ denote the estimated unconditional posterior mean and variance, respectively. The estimates of the two parameters $s_{1j}$ and $s_{0j}$ in  \eqref{gammapost} are
\begin{equation}\label{posttrueprob}
\hat{s}_{1j}=\frac{\hat{\mu}_{\lambda_j}^2}{\hat{\sigma}^2_{\lambda_j}},~\textrm{and}~~\hat{s}_{0j}=\frac{\hat{\mu}_{\lambda_j}}{\hat{\sigma}^2_{\lambda_j}}.
\end{equation}

To estimate the unconditional posterior moments of random effects in the Binomial model, we assume that hyper-parameters $r$  and  $\boldsymbol{\beta}$ are independent \emph{a posteriori}. With this assumption,  the unconditional posterior mean and variance of random effect $p_j$ are
\begin{align}
\E(p_{j}\mid \boldsymbol{y}) &=\E(\E(p_{j}\mid r, \boldsymbol{\beta}, \boldsymbol{y})\mid\boldsymbol{y})=(1-\E(B_j\mid\boldsymbol{y}))\bar{y}_j + \E(B_j\mid  \boldsymbol{y})\E(p^E_{j}\mid  \boldsymbol{y}),\label{postmean_binom} \\
~~~~~~\VAR(p_{j}\mid \boldsymbol{y}) &=  \E(\VAR(p_{j}\mid r, \boldsymbol{\beta}, \boldsymbol{y})\mid \boldsymbol{y})+\VAR(\E(p_{j}\mid r, \boldsymbol{\beta},\boldsymbol{y})\mid \boldsymbol{y})\label{postvar}\\
&= \E\big(p^{\ast}_{j}(1-p^{\ast}_{j})/ (r+n_{j}+1)\mid  \boldsymbol{y}\big)+\VAR\big(B_{j}(\bar{y}_{j}-p^E_{j})\mid \boldsymbol{y}\big) \\
&\approx \E\big(p^{\ast}_{j}(1-p^{\ast}_{j})(1-B_{j})/n_{i}\mid \boldsymbol{y}\big)+\VAR\big(B_{j}(\bar{y}_{j}-p^E_{j})\mid \boldsymbol{y}\big)\label{taylor}\\
&=\big\{(1-\bar{y}_j)\bar{y}_j[1-\E(B_j\mid \boldsymbol{y})]+(2\bar{y}_j-1)\E(B_j(1-B_j) \mid \boldsymbol{y})(\bar{y}_j-\E(p^E_j\mid\boldsymbol{y}))\nonumber\\
&~~~~+\E(B_j^2(1-B_j) \mid \boldsymbol{y})\E((\bar{y}_j-p^E_j)^2\mid\boldsymbol{y})\big\}/n_j+\VAR\big(B_{j}(\bar{y}_{j}-p^E_{j})\vert \boldsymbol{y}\big),\label{approximate_var}
\end{align}
where the approximation in \eqref{taylor} is a first-order Taylor approximation. By inserting the estimated unconditional posterior moments of shrinkage factors in \eqref{shrinkagemoments} and those of expected random effect in \eqref{est_exp_randomeffect} into  both \eqref{postmean_binom} and \eqref{approximate_var}, we obtain the estimates of the unconditional posterior mean and variance of each random effect, denoted by $\hat{\mu}_{p_j}$ and $\hat{\sigma}^2_{p_j}$, respectively. We thus obtain the estimates of two parameters $t_{1j}$ and $t_{0j}$ in  \eqref{betapost} as follows.
\begin{equation}\label{posttrueprob}
\hat{t}_{1j}=\left(\frac{\hat{\mu}_{p_j}(1-\hat{\mu}_{p_j})}{\hat{\sigma}^2_{p_j}}-1\right)\hat{\mu}_{p_j},~\textrm{and}~~\hat{t}_{0j}=\left(\frac{\hat{\mu}_{p_j}(1-\hat{\mu}_{p_j})}{\hat{\sigma}^2_{p_j}}-1\right)(1-\hat{\mu}_{p_j}).
\end{equation}

Finally, the assumed unconditional posterior distribution of random effect for the Gaussian model is
\begin{equation}\label{trueapprox_normal}
\mu_j\mid \boldsymbol{y} \sim \textrm{skewed-Normal}(\hat{\phi}, \hat{\omega}, \hat{\delta}),
\end{equation}
that for the Poisson model is
\begin{equation}\label{trueapprox_poisson}
\lambda_j\mid \boldsymbol{y} \sim \textrm{Gamma}(\hat{s}_{1j}, \hat{s}_{0j}).
\end{equation}
and that for the Binomial model is
\begin{equation}\label{trueapprox_binomial}
p_j\mid \boldsymbol{y} \sim \textrm{Beta}(\hat{t}_{1j}, \hat{t}_{0j}),
\end{equation}
Our point and interval estimates of each random effect are the mean and (2.5\%, 97.5\%) quantiles (if we assign 95\% confidence level) of the assumed unconditional posterior distribution in  \eqref{trueapprox_normal}, \eqref{trueapprox_poisson}, or \eqref{trueapprox_binomial}.

For the Binomial model,  \pkg{Rgbp} provides a fully Bayesian approach for drawing posterior samples of random effects and hyper-parameters, which we illustrate in the next section.

\section[Estimation]{The acceptance-rejection method for the Binomial model}\label{sec_accept}
In this section, we illustrate an option of  \pkg{Rgbp} that provides a way to draw  posterior samples of random effects and hyper-parameters via the acceptance-rejection (A-R) method \citep{robert2013monte} for the Binomial model. Unlike the approximate Bayesian machinery specified in the previous section, this method does not assume that hyper-parameters are independent \emph{a posteriori}. The joint posterior density function of $\alpha=-\log(r)$ and $\boldsymbol{\beta}$ based on their joint hyper-prior density function in  \eqref{eq:hyper} is 
\begin{equation}\label{ar_target}
f(\alpha, \boldsymbol{\beta} \mid \boldsymbol{y})\propto f(\alpha, \boldsymbol{\beta})L(\alpha, \boldsymbol{\beta})\propto \exp(\alpha)L(\alpha, \boldsymbol{\beta}).
\end{equation}

The A-R method  is useful when it is difficult to sample a parameter of interest $\theta$ directly from its target probability density $f(\theta)$, which is known up to a normalizing constant, but an easy-to-sample envelope function $g(\theta)$ is available. The A-R method samples $\theta$  from the envelope $g(\theta)$ and accepts it with a probability $f(\theta)/(Mg(\theta))$, where $M$ is a constant making $f(\theta)/g(\theta)\le M$ for all $\theta$. The distribution of the accepted $\theta$ exactly follows $f(\theta)$. The A-R method is stable as long as the tails of the envelope function are thicker than those of the target density function. 

The goal of the A-R method  for the  Binomial model is to  draw  posterior samples of hyper-parameters from \eqref{ar_target}, using an easy-to-sample envelope function $g(\alpha, \boldsymbol{\beta})$ that has thicker tails than the target density function. 

We factor the envelope function into two parts, $g(\alpha, \boldsymbol{\beta})=g_1(\alpha)g_2(\boldsymbol{\beta})$ to model the tails of each function separately. We consider the tail behavior of the conditional posterior density function $f(\alpha \mid \boldsymbol{\beta},  \boldsymbol{y})$ to establish $g_1(\alpha)$; $f(\alpha \mid \boldsymbol{\beta},  \boldsymbol{y})$ behaves as $\exp(-\alpha(k-1))$ when $\alpha$ goes to $\infty$ and as $\exp(\alpha)$ when $\alpha$ goes to $-\infty$. It indicates that $f(\alpha \mid \boldsymbol{\beta},  \boldsymbol{y})$ is skewed to the left because the right tail touches the $x$-axis faster than the left tail does as long as $k>1$.  A skewed $t$-distribution  is a good candidate for $g_1(\alpha)$ because it behaves as a power law on both tails, leading to thicker tails than those of $f(\alpha \mid \boldsymbol{\beta},  \boldsymbol{y})$. 

It is too complicated to figure out the tail behaviors of $f(\boldsymbol{\beta} \mid \alpha,  \boldsymbol{y})$. However, because $f(\boldsymbol{\beta} \mid \alpha,  \boldsymbol{y})$ in the Gaussian  model (as an approximation) has a multivariate Gaussian density function \citep{tang2011, kelly2014advances}, we consider a multivariate $t$-distribution with four  degrees of freedom as a good candidate for $g_2(\boldsymbol{\beta})$.

Specifically, we assume
\begin{eqnarray}\label{env}
g_1(\alpha) &=& g_1(\alpha; l, \sigma, a, b)~\equiv~\textrm{Skewed-}t(\alpha\mid l, \sigma, a, b),\\
g_2(\boldsymbol{\beta}) &=& g_2(\boldsymbol{\beta}; \boldsymbol{\xi}, S_{(m\times m)})~\equiv~t_{4}(\boldsymbol{\beta}\mid \boldsymbol{\xi}, S),\label{g2approx}
\end{eqnarray}
where Skewed-$t(\alpha\mid l, \sigma, a, b)$ represents a density function of a skewed $t$-distribution  of $\alpha$ with location $l$, scale $\sigma$, degree of freedom $a+b$, and skewness $a-b$ for any positive constants $a$ and $b$ \citep{jones2003skew}. \cite{jones2003skew} derive the mode of $g_1(\alpha)$ as
\begin{equation}\label{mode}
l+\frac{(a-b)\sqrt{a+b}}{\sqrt{(2a+1)(2b+1)}},
\end{equation} 
and provide a  representation to generate  random variables that follows Skewed-$t(\alpha\mid l, \sigma, a, b)$;
\begin{equation}
\alpha\sim l+\sigma\frac{\sqrt{a+b}(2T-1)}{2\sqrt{T(1-T)}}, ~\textrm{where}~ T\sim \textrm{Beta}(a, b).
\end{equation}

They also show that the tails of the skewed-$t$ density function  follow a power law with $\alpha^{-(2a+1)}$ on the left and $\alpha^{-(2b+1)}$ on the right when $b>a$. 

The notation $t_{4}(\boldsymbol{\beta}\mid\boldsymbol{\xi}, S)$ in  \eqref{g2approx} indicates a density function of a multivariate $t$-distribution of $\boldsymbol{\beta}$ with four degrees of freedom, a location vector $\boldsymbol{\xi}$, and a $m\times m$ scale matrix $S$ that leads to the variance-covariance matrix $2S$. 

\pkg{Rgbp} determines the parameters of $g_1(\alpha)$ and $g_2(\boldsymbol{\beta})$, \emph{i.e.}, $l$, $\sigma$, $a$, $b$, $\boldsymbol{\xi}$, and $S$, to make the product of $g_1(\alpha)$ and $g_2(\boldsymbol{\beta})$ similar to the target joint posterior density $f(\alpha, \boldsymbol{\beta} \mid  \boldsymbol{y})$. First, \pkg{Rgbp} obtains the  mode of $f(\alpha, \boldsymbol{\beta} \mid  \boldsymbol{y})$, $(\hat{\alpha}, \hat{\boldsymbol{\beta}})$, and the inverse of the  negative Hessian matrix at the mode. We define $-H^{-1}_{\hat{\alpha}}$ to indicate the  (1, 1)$^{\textrm{th}}$ element of the negative Hessian matrix and $-H^{-1}_{\hat{\boldsymbol{\beta}}}$ to represent  the negative Hessian matrix without the first row and  the first column.  

For $g_1(\alpha)$, \pkg{Rgbp} sets ($a,~ b$) to $(k,~ 2k)$ if $k$ is less than 10 (or  to $(\log(k),~ 2\log(k))$ otherwise)   for a left-skewness  and  these small values of $a$ and $b$ lead to thick tails. \pkg{Rgbp} matches the mode of $g_1(\alpha)$ specified in  \eqref{mode} to $\hat{\alpha}$ by setting the location parameter $l$ to $\hat{\alpha}-(a-b)\sqrt{a+b}/\sqrt{(2a+1)(2b+1)}$. \pkg{Rgbp} sets  the scale parameter $\sigma$ to $(-H^{-1}_{\hat{\alpha}})^{0.5}\psi$, where $\psi$ is a tuning parameter; when the A-R method produces extreme weights defined in  \eqref{weight} below, we need enlarge the value of $\psi$.  

For $g_2(\boldsymbol{\beta})$,  \pkg{Rgbp} sets the location vector $\boldsymbol{\xi}$ to the mode $\hat{\boldsymbol{\beta}}$ and the scale matrix $S$ to $-H^{-1}_{\hat{\boldsymbol{\beta}}}/2$ so that the variance-covariance matrix becomes $-H^{-1}_{\hat{\boldsymbol{\beta}}}$.

For  implementation of the acceptance-rejection method, \pkg{Rgbp} draws four times more trial samples than the desired number of samples, denoted by $N$, independently from $g_1(\alpha)$ and $g_2(\boldsymbol{\beta})$. \pkg{Rgbp} calculates $4N$ weights, each of which is defined as
\begin{equation}\label{weight}
w_i\equiv w(\alpha^{(i)}, \boldsymbol{\beta}^{(i)})=\frac{f(\alpha^{(i)}, \boldsymbol{\beta}^{(i)}\mid \boldsymbol{y})}{g_1(\alpha^{(i)})g_2( \boldsymbol{\beta}^{(i)})}, ~\textrm{for}~ i=1, 2, \ldots, 4N.
\end{equation}
\pkg{Rgbp} accepts each pair of $(\alpha^{(i)}, \boldsymbol{\beta}^{(i)})$ with a probability $w_i/M$ where $M$ is set to the maximum of all the $4N$ weights. When \pkg{Rgbp} accepts more than  $N$ pairs, it discards the redundant. If \pkg{Rgbp} accepts less than $N$ pairs, then it additionally draws   $N'$ (six times the shortage) pairs    and calculates a new maximum $M'$ from all the previous and new weights; \pkg{Rgbp} accepts or rejects the entire pairs again with new probabilities $w_j/M'$, $j=1, 2, \ldots, 4N+N'$.

After obtaining posterior samples of hyper-parameters, \pkg{Rgbp} draws posterior samples of random effects from $f(\boldsymbol{p}\mid \boldsymbol{y})$ in \eqref{mcintegration_binomial}. The integration on the right hand side of \eqref{mcintegration_binomial} can be done by sampling $\boldsymbol{p}$ from $f(p_j\mid \boldsymbol{\beta}, r, \boldsymbol{y})$ in  \eqref{beta_post} for $j=1, 2, \ldots, k$,  given $r=\exp(-\alpha)$ and $\boldsymbol{\beta}$ that are already sampled from $f(\alpha, \boldsymbol{\beta}\mid \boldsymbol{y})$ via the A-R method.

\section{Frequency method checking}\label{sec4}
The question as to whether the interval estimates of random effects for given confidence level obtained by a specific model achieve the nominal coverage rate  for any true parameter values is one of the key model evaluation criteria. Unlike standard model checking methods that test whether a two-level model is appropriate for data  \citep{dean1992testing, modelchecking1996}, frequency method checking is a procedure to evaluate the coverage properties of the model.   Conditioning that the two-level model is appropriate, the frequency method checking generates pseudo-data sets given specific values of hyper-parameters  and estimates unknown coverage probabilities based on these mock data sets (a parametric bootstrapping). We  describe the frequency method checking based on the Gaussian model because the idea can be easily applied to the other two models.

\subsection{Pseudo-data generation}\label{data_generation}
\begin{figure}[h!]
\begin{center}
\includegraphics[width=5cm]{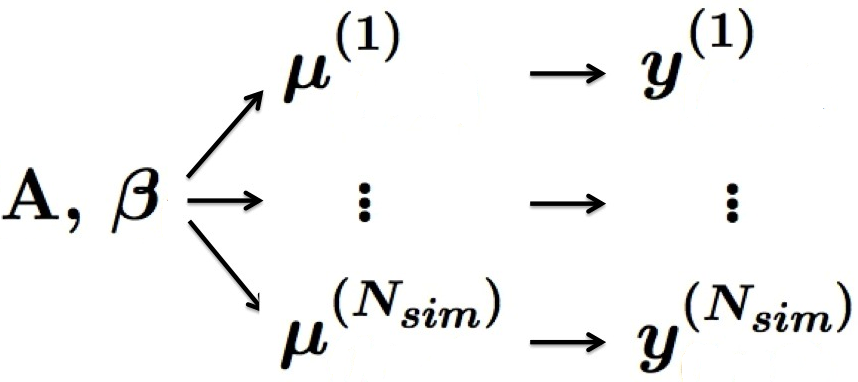}
\caption{Pseudo-data generating process.}
\label{fig:pseudo}
\end{center}
\end{figure}

Figure \ref{fig:pseudo} displays the process of generating pseudo-data sets. It is noted that  the conjugate prior distribution of each random effect in  \eqref{normalprior} is completely determined by two hyper-parameters, $A$ and $\boldsymbol{\beta}$. Fixing these hyper-parameters at specific values, we generate  $N_{\textrm{sim}}$ sets of random effects from the conjugate prior distribution, \emph{i.e.}, \{{\boldmath $\mu$}$^{(i)},~i=1, \ldots, N_{\textrm{sim}}\}$, where the superscript $(i)$ indicates the $i$-th simulation. Next, using the distribution of observed data in  \eqref{normalobs}, we generate $N_{\textrm{sim}}$ sets of observed data sets $\{\boldsymbol{y}^{(i)},~i=1, \ldots, N_{\textrm{sim}}\}$ given each {\boldmath$\mu$}$^{(i)}$. 

\subsection{Coverage probability estimation}
After fitting the Gaussian model for each simulated data set, we  obtain interval estimates of the random effects $\boldsymbol{\mu}^{(i)}$. Let $(\hat{\mu}^{(i)}_{j, ~\textrm{low}}, ~\hat{\mu}^{(i)}_{j, ~\textrm{upp}})$ represent the lower and upper bounds of the interval estimate of random effect $j$ based on the $i$-th simulation given a specific confidence level.  We define the coverage indicator of random effect $j$ on the $i$-th mock data set as 
\begin{equation}\label{coverage_indicator}
I\left(\mu_j^{(i)}\right) = \left\{ \begin{array}{ll}
1, & \textrm{if $\mu_j^{(i)}\in\left(\hat{\mu}^{(i)}_{j, ~\textrm{low}}, ~\hat{\mu}^{(i)}_{j, ~\textrm{upp}}\right)$,}\\
0, & \textrm{otherwise.}
\end{array} \right.
\end{equation}
This shrinkage indicator is equal to the value one if the random effect $j$ in simulation $i$ is between its interval estimates and zero otherwise.





\subsubsection{Simple unbiased coverage estimator.}
When the confidence level is 95\%, the proportion of 95\% interval estimates that contain random effect $j$ is an intuitive choice for  the coverage rate estimator for random effect $j$. This estimator  implicitly assumes that there exist $k$ unknown coverage probabilities of random effects, denoted by $C_{A, \boldsymbol{\beta}}(\mu_j)$ for $j=1, 2, \ldots, k$, depending on the values of the hyper-parameters that generate random effects and mock data sets. The coverage indicators for random effect $j$ in  \eqref{coverage_indicator} is assumed to follow an independent and identically distributed  Bernoulli distribution given the unknown coverage rate $C_{A, \boldsymbol{\beta}}(\mu_j)$. The sample mean of these coverage indicators is a simple unbiased coverage estimator for $C_{A, \boldsymbol{\beta}}(\mu_j)$;  for $j=1, 2, \ldots, k$,
\begin{equation}
\bar{I}(\mu_j)= \frac{1}{N_{\textrm{sim}}}\sum_{i=1}^{N_{\textrm{sim}}}I\left(\mu_j^{(i)}\right).
\end{equation}
The unbiased variance estimator of $\VAR(\bar{I}(\mu_j))$ is, for $j=1, 2, \ldots, k$,
\begin{equation}\label{svar}
\widehat{\VAR}\left(\bar{I}(\mu_j)\right)=\frac{1}{N_{\textrm{sim}}(N_{\textrm{sim}}-1)}\sum_{i=1}^{N_{\textrm{sim}}}\left(I(\mu_j^{(i)})-\bar{I}(\mu_j)\right)^{2}.
\end{equation}

\subsubsection{Rao-Blackwellized unbiased coverage estimator.}
Frequency method checking is computationally expensive in nature because it  fits a model on every mock data set. The situation deteriorates if the number of simulations or the size of data is large, or the estimation method is computationally demanding. \citet{morris1997} and \cite{tang2002fitting} use a Rao-Blackwellized (RB) unbiased coverage estimator for the unknown coverage rate of each random effect, which is more efficient than the simple unbiased coverage estimator.  For $j=1, 2, \ldots, k$,
\begin{equation}\label{RB_theory}
C_{A, \boldsymbol{\beta}}(\mu_j)=\E\!\left(\bar{I}(\mu_j)\mid A, \boldsymbol{\beta}\right)=\E\bigg[\frac{1}{N_{\textrm{sim}}}\sum_{i=1}^{N_{\textrm{sim}}}\E\!\left(I(\mu_j^{(i)})\mid A, \boldsymbol{\beta}, \boldsymbol{y}^{(i)}\right)\bigg\vert~ A, \boldsymbol{\beta}\bigg],
\end{equation}
where the sample mean of the interior conditional expectations in \eqref{RB_theory} is the RB unbiased coverage estimator.  Specifically,
\begin{align}
\bar{I}^{RB}(\mu_j) &= \frac{1}{N_{\textrm{sim}}}\sum_{i=1}^{N_{\textrm{sim}}}\E\!\left(I(\mu_j^{(i)})\mid A, \boldsymbol{\beta}, \boldsymbol{y}^{(i)}\right)\label{RB1}\\
&= \frac{1}{N_{\textrm{sim}}}\sum_{i=1}^{N_{\textrm{sim}}} \Prob\!\left(\mu_j^{(i)}\in(\hat{\mu}^{(i)}_{j, ~\textrm{low}}, ~\hat{\mu}^{(i)}_{j, ~upp})\mid A, \boldsymbol{\beta},  \boldsymbol{y}^{(i)}\right)\!.\label{RB2}
\end{align}
We  can easily compute the  conditional posterior probabilities in \eqref{RB2} using the cumulative density function of the Gaussian conditional posterior distribution of each random effect in  \eqref{normalpost}. The variance of  $\bar{I}^{RB}(\mu_j)$ does not exceed the variance of a simple unbiased coverage estimator, $\bar{I}(\mu_j)$ \citep{radhakrishna1945information, blackwell1947conditional}.


If one dataset $\boldsymbol{y}^{(i)}$ is simulated for each set of random effects $\boldsymbol{\mu}^{(i)}$, the variance estimator below is an unbiased estimator of $\VAR(\bar{I}^{RB}(\mu_j) )$. For $j=1, 2, \ldots, k,$
\begin{equation}\label{RBvar}
\widehat{\VAR}(\bar{I}^{RB}(\mu_j) )\equiv\frac{1}{N_{\textrm{sim}}(N_{\textrm{sim}}-1)}\sum_{i=1}^{N_{\textrm{sim}}}\left(\E(I(\mu_j^{(i)})\mid A, \boldsymbol{\beta}, \boldsymbol{y}^{(i)})-\bar{I}^{RB}(\mu_j) \right)^{2}.
\end{equation}

\subsubsection{Overall unbiased coverage estimator}\label{overallRB} To summarize the frequency method checking, we report the overall unbiased coverage estimate and its variance estimate,
\begin{equation}\label{RBoverall}
\bar{\bar{I}}^{RB} = \frac{1}{k}\sum_{j=1}^k\bar{I}^{RB}(\mu_j)~~\textrm{and}~~ \widehat{\VAR}(\bar{\bar{I}}_{RB})=\frac{1}{k^2}\sum_{j=1}^k\widehat{\VAR}(\bar{I}^{RB}(\mu_j)).
\end{equation}

\section[Examples]{Examples}\label{sec6}
In this section, we demonstrate how \pkg{Rgbp} can be used to analyze three realistic data sets:  Medical profiling of 31 hospitals with Poisson distributed fatality counts; educational assessment of eight schools with Normally distributed data; and evaluation of 18 baseball hitters with Binomial success rates and one covariate. For each example, we construct 95\% confidence intervals. Additional usages and options of the functions in \pkg{Rgbp} can be found in Section   \ref{sec5}.

\subsection[Known Second-level Mean]{Poisson data with 31 hospitals: Known expected random effect}
\label{sec:ex:hosp}


We analyze a data set of 31 hospitals in New York State consisting of the outcomes of the coronary artery bypass graft (CABG) surgery \citep{morris2012}. The data set contains the number of deaths, $\boldsymbol{y}$, for a specified period after CABG surgeries out of the total number of patients, $\boldsymbol{n}$, receiving CABG surgeries in each hospital. A goal would be to obtain the point and interval estimates for the unknown true fatality rates (random effects) of 31 hospitals to evaluate  each hospital's reliability on the CABG surgery (\cite{morris1995} use a similar Poisson model to handle these hospital profile data). We interpret the caseloads, $\boldsymbol{n}$,  as exposures and assume that the state-level fatality rate per exposure of this surgery  is known, $\lambda^E_j= 0.03$ ($m=0$). 


The following code can be used to load these data into \proglang{R}.
\begin{CodeChunk}
\begin{CodeInput}
R> library("Rgbp")
R> data("hospital")
R> y <- hospital$d
R> n <- hospital$n
\end{CodeInput}
\end{CodeChunk}



The function \code{gbp} can then be used to fit a Poisson-Gamma to the fatality
rates in New York States with the expected random effect, $\lambda^E_j$, equal to 0.03.

\begin{CodeChunk}
\begin{CodeInput}
R> p.output <- gbp(y, n, mean.PriorDist = 0.03, model = "poisson")
R> p.output
\end{CodeInput}
\begin{CodeOutput}
Summary for each unit (sorted by n):

      obs.mean    n prior.mean shrinkage low.intv post.mean upp.intv post.sd
1       0.0448   67       0.03     0.911   0.0199    0.0313   0.0454 0.00653
2       0.0294   68       0.03     0.910   0.0189    0.0299   0.0435 0.00631
3       0.0238  210       0.03     0.765   0.0185    0.0285   0.0407 0.00566
4       0.0430  256       0.03     0.728   0.0225    0.0335   0.0467 0.00619
5       0.0335  269       0.03     0.718   0.0208    0.0310   0.0432 0.00573
6       0.0438  274       0.03     0.714   0.0229    0.0339   0.0472 0.00621
7       0.0432  278       0.03     0.711   0.0228    0.0338   0.0469 0.00617
8       0.0136  295       0.03     0.699   0.0157    0.0250   0.0366 0.00534
9       0.0288  347       0.03     0.663   0.0200    0.0296   0.0410 0.00536
10      0.0372  349       0.03     0.662   0.0222    0.0325   0.0446 0.00571
11      0.0391  358       0.03     0.656   0.0228    0.0331   0.0454 0.00579
12      0.0177  396       0.03     0.633   0.0165    0.0255   0.0363 0.00506
13      0.0278  431       0.03     0.613   0.0200    0.0292   0.0400 0.00511
14      0.0249  441       0.03     0.608   0.0191    0.0280   0.0387 0.00502
15      0.0273  477       0.03     0.589   0.0199    0.0289   0.0394 0.00499
16      0.0455  484       0.03     0.585   0.0256    0.0364   0.0491 0.00601
17      0.0304  494       0.03     0.580   0.0211    0.0302   0.0409 0.00506
18      0.0220  501       0.03     0.577   0.0180    0.0266   0.0369 0.00483
19      0.0277  505       0.03     0.575   0.0202    0.0290   0.0395 0.00494
20      0.0204  540       0.03     0.559   0.0173    0.0258   0.0358 0.00474
21      0.0284  563       0.03     0.548   0.0206    0.0293   0.0395 0.00485
22      0.0236  593       0.03     0.535   0.0187    0.0270   0.0369 0.00466
23      0.0150  602       0.03     0.532   0.0147    0.0230   0.0329 0.00466
24      0.0238  629       0.03     0.521   0.0188    0.0271   0.0368 0.00460
25      0.0204  636       0.03     0.518   0.0173    0.0254   0.0351 0.00455
26      0.0480  729       0.03     0.484   0.0286    0.0393   0.0516 0.00587
27      0.0306  849       0.03     0.446   0.0223    0.0303   0.0397 0.00445
28      0.0274  914       0.03     0.428   0.0208    0.0285   0.0374 0.00423
29      0.0213  940       0.03     0.421   0.0176    0.0249   0.0335 0.00407
30      0.0293 1193       0.03     0.364   0.0223    0.0296   0.0379 0.00397
31      0.0201 1340       0.03     0.338   0.0170    0.0235   0.0310 0.00360
Mean            517       0.03     0.600   0.0201    0.0293   0.0403 0.00517
\end{CodeOutput}
\end{CodeChunk}

The output contains information about  (from the left) the observed fatality rates $\bar{y}_{j}$, caseloads $n_{j}$, known expected random effect $\lambda^E_j$, shrinkage estimates $\hat{B}'_{j}$, lower bounds (2.5\%) of posterior interval estimates $\hat{\lambda}_{j, \textrm{low}}$, posterior means $\hat{\E}(\lambda_j\vert\boldsymbol{y})$, upper bounds  (97.5\%) of posterior interval estimates $\hat{\lambda}_{j, \textrm{upp}}$, and posterior standard deviations $\widehat{\textrm{SD}}(\lambda_j\vert\boldsymbol{y})$ for  random effects based on the assumed unconditional Gamma posterior distributions in \eqref{trueapprox_poisson}. 


A function \code{summary} shows selective information about hospitals with minimum, median, and maximum exposures and the estimation result of the hyper-parameter $\alpha=-\log(r)$.  \begin{CodeChunk}
\begin{CodeInput}
R> summary(p.output)
\end{CodeInput}
\begin{CodeOutput}
Main summary:

                    obs.mean    n prior.mean shrinkage low.intv post.mean
Unit with min(n)      0.0448   67       0.03     0.911   0.0199    0.0313   
Unit with median(n)   0.0455  484       0.03     0.585   0.0256    0.0364   
Unit with max(n)      0.0201 1340       0.03     0.338   0.0170    0.0235   
Overall Mean                  517       0.03     0.600   0.0201    0.0293   

                    upp.intv  post.sd
                      0.0454  0.00653
                      0.0491  0.00601
                      0.0310  0.00360
                      0.0403  0.00517

Second-level Variance Component Estimation Summary:
alpha=log(A) for Gaussian or alpha=log(1/r) for Binomial and Poisson data:

post.mode.alpha post.sd.alpha post.mode.r
          -6.53         0.576         684
\end{CodeOutput}
\end{CodeChunk}
The output of \code{summary} shows that $\hat{r}=\textrm{exp}(6.53)=684$, which is an indicator of how valuable and informative the second-level hierarchy is. It means that the 25 hospitals with caseload less than 684 patients shrink their sample means towards the prior mean (0.03) more than 50\%. For example, the shrinkage estimate of the first hospital ($\hat{B}_{1}= 0.911$) was calculated by 684 / (684 + 67), where 67 is its caseload ($n_{1}$). As for this hospital, using more information from the conjugate prior distribution is an appropriate choice because the amount  of observed information (67) is much less than the amount of state-level information (684).


To obtain a graphical summary, we use the function \code{plot}.

\begin{CodeChunk}
\begin{CodeInput}
R> plot(p.output)
\end{CodeInput}
\end{CodeChunk}
\begin{figure}[h]
\begin{center}
\includegraphics[width = 2.7in]{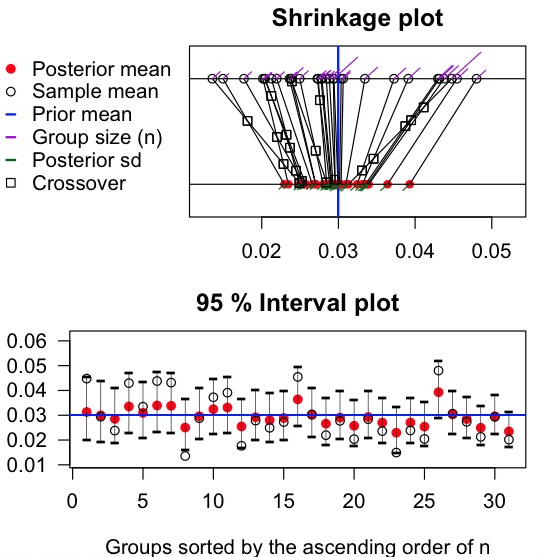}
\caption{Shrinkage plot and 95\% interval plot for fatality rates at 31 hospitals.}
\label{fig:hospshr}
\end{center}
\end{figure}

The shrinkage plot \citep{1975, morris2012} in the first panel of Figure~\ref{fig:hospshr} shows the regression towards the mean; the observed fatality rates, denoted by empty dots on the upper horizontal line, are shrinking towards the known expected random effect, denoted by a blue vertical line at 0.03, to the different extents. The red dots on the bottom line denotes the estimated posterior means. Some hospitals' ranks have changed by shrinking more sharply towards 0.03 than the others. For example, an empty square at the crossing point of the two left-most lines (8th and 23rd hospitals on the list above) indicates that  the seemingly safest hospital in terms of the observed mortality rate is probably not the safest in terms of the estimated posterior mean accounting for the different caseloads of these two hospitals. 

To be specific, their observed fatality rates ($y_{j}$, $j=8, 23$) are 0.0136 and 0.0150 and caseloads ($n_{j}$, $j=8, 23$) are 295 and 602, respectively. Considering solely the observed fatality rates may lead to an unfair comparison because the latter hospital handled twice the caseload. \pkg{Rgbp} accounts for this caseload difference, making the posterior mean for the random effect of the former hospital shrink toward the state-level mean ($\lambda^E_j$=0.03) more rapidly than that for the latter hospital.

The point estimates are not enough to evaluate hospital reliability because one hospital may have a lower point estimate but  larger uncertainty (variance) than the other. The second plot of Figure \ref{fig:hospshr} displays the 95\% interval estimates. Each posterior mean (red dot) is between the sample mean (empty dot) and the known expected random effect (a blue horizontal line). 

This 95\% interval plot reveals that the 31st hospital has the lowest upper bound even though its point estimate ($\hat{\lambda}_{31}=0.0235$) is slightly larger than that of the 23rd hospital ($\hat{\lambda}_{23}=0.0230$). The observed  mortality rates for these two hospitals ($y_{j}, j=23, 31$) are 0.0150 and 0.0201 and the caseloads ($n_{j}, j =23, 31$) are 602 and 1340 each. The 31st hospital has twice the caseload, which leads to borrowing less information from the New York State-level hierarchy (or shrinking less toward the state-level mean, 0.03) with smaller variance. Based on the point and interval estimates, the 31st hospital seems a better choice than the 23rd hospital. (Note that this conclusion is based on the data, assuming no covariate information  about the overall case difficulties in each hospital. A more reliable analysis must take into account all the possible covariate information and instead of our Poisson model, we recommend using our Binomial model to account for covariate information.)

Next, we perform frequency method checking to question how reliable the estimation procedure is, assuming $r$ equals its estimated value, $\hat{r}=683.53$.  The function \code{coverage}  generates synthetic data sets starting with the estimated value of $r$  as a generative value. For reference, we could designate other generative values of $r$ and $\lambda^E_j$ by adding two arguments, \code{A.or.r} and \code{mean.PriorDist}, into the code below, see Section \ref{code_model} for details.


\begin{CodeChunk}
\begin{CodeInput}
R> p.coverage <- coverage(p.output, nsim = 1000)
\end{CodeInput}
\end{CodeChunk}
\begin{figure}[h] 
\begin{center}
\includegraphics[width = 2.7in]{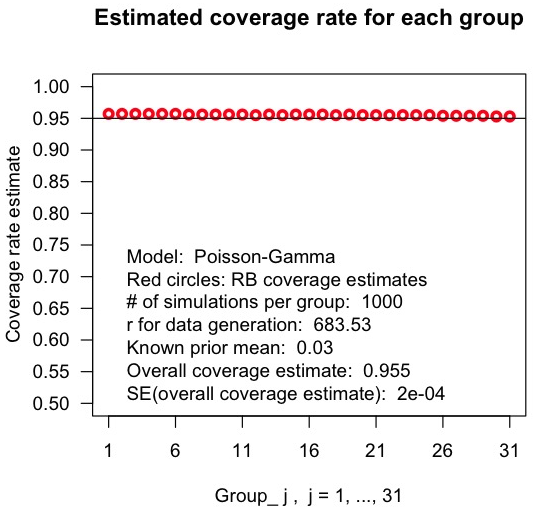}
\caption{Coverage plot via frequency method checking for 31 hospitals.}
\label{fig:hospitalcoverage}
\end{center}
\end{figure}

In Figure \ref{fig:hospitalcoverage}, the black horizontal line at 0.95 represents the nominal confidence level  and the red circles indicate RB unbiased coverage estimates, $\bar{I}^{RB}(\lambda_j)$ for $j=1, 2, \ldots, 31$. The  overall unbiased coverage estimate across all the hospitals, $\bar{\bar{I}}^{RB}$ in \eqref{RBoverall}, is 0.955. None of RB unbiased coverage estimates for the 31 hospitals are less than 0.95 regardless of their caseloads, which range from 67 for hospital 1 to 1,340 for hospital 31. This result shows that the interval estimates for this particular dataset adequately achieves a 95\% confidence level if $r=\hat{r}$.


The following code provides 31 RB unbiased coverage estimates and their standard errors  (the output is omitted for space reasons).
\begin{CodeChunk}
\begin{CodeInput}
R> p.coverage$coverageRB
R> p.coverage$se.coverageRB
\end{CodeInput}
%\begin{CodeOutput}
% [1] 0.955 0.954 0.955 0.954 0.954 0.954 0.954 0.954 0.954 0.954 0.954 0.954
%[13] 0.954 0.953 0.953 0.954 0.953 0.954 0.954 0.953 0.953 0.953 0.954 0.953
%[25] 0.953 0.953 0.952 0.953 0.952 0.952 0.952
%\end{CodeOutput}
\end{CodeChunk}

The code below produces 31 simple unbiased coverage estimates and their standard errors.
\begin{CodeChunk}
\begin{CodeInput}
R> p.coverage$coverageS
R> p.coverage$se.coverageS
\end{CodeInput}
%\begin{CodeOutput}
% [1] 0.949 0.960 0.958 0.958 0.950 0.945 0.960 0.953 0.960 0.956 0.955 0.946 
%[13] 0.955 0.954 0.960 0.965 0.955 0.952 0.960 0.956 0.959 0.955 0.964 0.960 
%[25] 0.945 0.942 0.947 0.960 0.940 0.946 0.956
%\end{CodeOutput}
\end{CodeChunk}



It turns out that the variance estimate of the RB unbiased coverage estimate for the first hospital ($0.0016^2$) is about 19 times smaller than that of the simple one ($0.0070^2$). It means that the RB unbiased coverage estimates based on 1,000 simulations ($N_{\textrm{sim}}$) are as precise as the simple unbiased coverage estimates based on 19,000 simulations in terms of estimating the coverage probability for the first hospital, $C_{r, \lambda^E}(\lambda_1)$.



\subsection[Unknown Second-level Mean and No Covariate]{Gaussian data with eight schools: Unknown expected random effect and no covariates} \label{sec:ex:8schools}

The Education Testing Service conducted randomized experiments in eight
separate schools (groups) to test whether students (units) SAT scores are
affected by coaching. The dataset contains the estimated coaching effects on SAT
scores ($y_{j},~ j=1, \ldots, 8$) and standard errors ($V^{0.5}_{j},~ j=1, \ldots, 8$)
of the eight schools \citep{1981}. These data are contained in the package and can be loaded into \proglang{R} as follows.
\begin{CodeChunk}
\begin{CodeInput}
R> library("Rgbp")
R> data("schools")
R> y  <- schools$y
R> se <- schools$se
\end{CodeInput}
\end{CodeChunk}

Due to the nature of the test each school's coaching effect has an approximately Normal sampling distribution with approximately known sampling variances, based on large sample consideration.  At the second hierarchy, the mean for each school is assumed to be drawn from a common Normal distribution ($m=1$).

\begin{CodeChunk}
\begin{CodeInput}
R> g.output <- gbp(y, se, model = "gaussian")
R> g.output
\end{CodeInput}
\begin{CodeOutput}
Summary for each group (sorted by the descending order of se): 

     obs.mean   se prior.mean shrinkage low.intv post.mean upp.intv post.sd
8       12.00 18.0       8.17     0.734   -10.21      9.19     29.9   10.23
3       -3.00 16.0       8.17     0.685   -17.13      4.65     22.5   10.10
1       28.00 15.0       8.17     0.657    -2.32     14.98     38.8   10.56
4        7.00 11.0       8.17     0.507    -8.78      7.59     23.6    8.26
6        1.00 11.0       8.17     0.507   -13.03      4.63     20.1    8.44
2        8.00 10.0       8.17     0.459    -7.25      8.08     23.4    7.81
7       18.00 10.0       8.17     0.459    -1.29     13.48     30.8    8.18
5       -1.00  9.0       8.17     0.408   -13.30      2.74     16.7    7.63
Mean          12.5       8.17     0.552    -9.16      8.17     25.7    8.90
\end{CodeOutput}
\end{CodeChunk}
This output from \code{gbp} summarizes the results. In this Gaussian  model the amount of shrinkage for each unit is governed by the shrinkage factor, $B_j = V_j/(V_j + A)$. As such, schools whose variation within the school ($V_{j}$) is less than the between-school variation ($A$) will shrink greater than $50\%$. The results provided by \code{gpb} suggests that there is little evidence that the training provided much added benefit due to the fact that every school's $95\%$ posterior interval contains zero. In the case where the number of groups is large \pkg{Rgbp} provides a summary feature:

\begin{CodeChunk}
\begin{CodeInput}
R> summary(g.output)
\end{CodeInput}
\begin{CodeOutput}
Main summary:

                      obs.mean   se prior.mean shrinkage low.intv post.mean
Unit with min(se)        -1.00  9.0       8.17     0.408   -13.30      2.74     
Unit with median(se)1     1.00 11.0       8.17     0.507   -13.03      4.63     
Unit with median(se)2     7.00 11.0       8.17     0.507    -8.78      7.59     
Unit with max(se)        12.00 18.0       8.17     0.734   -10.21      9.19     
Overall Mean                   12.5       8.17     0.552    -9.16      8.17     

                      upp.intv post.sd
                          16.7    7.63
                          20.1    8.44
                          23.6    8.26
                          29.9   10.23
                          25.7    8.90

Second-level Variance Component Estimation Summary:
alpha=log(A) for Gaussian or alpha=log(1/r) for Binomial and Poisson data:

post.mode.alpha post.sd.alpha post.mode.A
           4.77          1.14         118


Regression Summary:

      estimate   se z.val p.val
beta1    8.168 5.73 1.425 0.154
\end{CodeOutput}
\end{CodeChunk}
The summary provides results regarding the second level hierarchy parameters. It can be seen that the estimate of the expected random effect, $\mu^E=\beta_1$ (\code{beta1}), is not significantly different from zero suggesting that there is no effect of the coaching program on SAT math scores.

\begin{figure}[h] 
\begin{center}
\includegraphics[width = 2.7in]{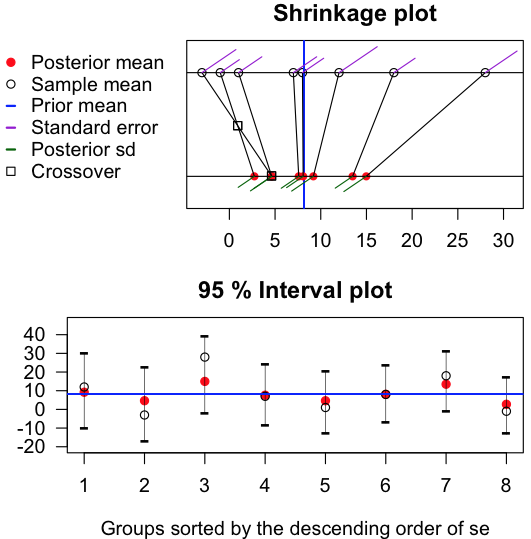}
\caption{Shrinkage plot and 95\% interval plot for eight schools.}
\label{fig:8schoolsplot}
\end{center}
\end{figure}

\pkg{Rgbp} also provides functionality to plot the results of the analysis as seen in Figure \ref{fig:8schoolsplot}. Plotting the results provides a visual aid to understanding but is only largely beneficial when the number of groups $(k)$ is small. 

\begin{CodeChunk}
\begin{CodeInput}
R> plot(g.output)
\end{CodeInput}
\end{CodeChunk}

The frequency method checking generates new pseudo-data from our assumed model. Unless otherwise specified, the procedure fixes the hyper-parameter values at their estimates ($\hat{A}$ and $\hat{\boldsymbol{\beta}}_1$ in this example) and then simulates random effects $\mu_j$ for each group $j$. The model is then estimated and this is repeated an $N_{\textrm{sim}}$ (\code{nsim}) number of times to estimate the coverage probabilities of the procedure.  

\begin{CodeChunk}
\begin{CodeInput}
R> g.coverage <- coverage(g.output, nsim = 1000)
\end{CodeInput}
\end{CodeChunk}
\begin{figure}[h!] 
\begin{center}
\includegraphics[width = 2.7in]{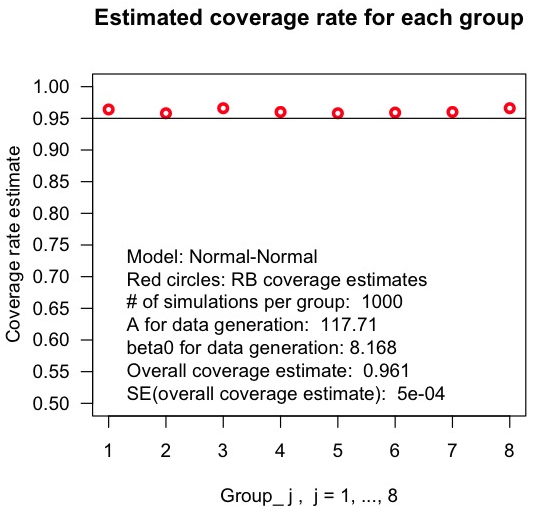}
\caption{Coverage plot via frequency method checking for 8 schools.}
\label{fig:schoolcoverage}
\end{center}
\end{figure}

As seen in Figure \ref{fig:schoolcoverage} the desired $95\%$ confidence level, denoted by a black horizontal line at 0.95, is achieved for each school in this example. All the coverage estimates depend on the chosen generative values of $A$ and $\beta_{1}$, and the assumption that the model is valid.

In addition, RB unbiased coverage estimate and its standard error for each school can be calculated with the command below.
\begin{CodeChunk}
\begin{CodeInput}
R> g.coverage$coverageRB
\end{CodeInput}
\begin{CodeOutput}
 [1] 0.966 0.959 0.967 0.960 0.959 0.962 0.960 0.966
\end{CodeOutput}
\end{CodeChunk}
\begin{CodeChunk}
\begin{CodeInput}
R> g.coverage$se.coverageRB
\end{CodeInput}
\begin{CodeOutput}
 [1] 0.0013 0.0012 0.0013 0.0013 0.0011 0.0011 0.0010 0.0017
\end{CodeOutput}
\end{CodeChunk}




\subsection[Unknown Second-level Mean and One Covariate]{Binomial data with 18 baseball players:  Unknown expected random effects and one covariate} 

The data of 18 major league baseball players contain the batting averages
through their first 45 official at-bats of the 1970 season \citep{1975}. A
binary covariate is created that is equal to the value one if a player is an
outfielder and zero otherwise. The data can be loaded into \proglang{R} with the following code.
\begin{CodeChunk}
\begin{CodeInput}
R> library("Rgbp")
R> data("baseball")
R> y <- baseball$Hits
R> n <- baseball$At.Bats
R> x <- ifelse(baseball$Position == "fielder", 1, 0)
\end{CodeInput}
\end{CodeChunk}
Conditional on the unknown true batting average (random effect) of each player
it is assumed that the at-bats are independent and therefore, $y_{j}\mid p_{j}\sim\textrm{Binomial}(45,~ p_{j})$ independently for $j=1, \ldots, 18$. Our goal is to obtain point and interval estimates of each random effect whilst considering the additional information on whether the player is an outfielder or not. The function \code{gbp} provides a way to incorporate such covariate information seamlessly into the model so that the regression towards the mean occurs within outfielders and non-outfielders separately. 

\begin{CodeChunk}
\begin{CodeInput}
R> b.output <- gbp(y, n, x, model = "binomial")
R> b.output
\end{CodeInput}
\begin{CodeOutput}
Summary for each unit (sorted by n):

     obs.mean  n  X1 prior.mean shrinkage low.intv post.mean upp.intv post.sd
1       0.400 45 1.0      0.310     0.715    0.248     0.335    0.429  0.0462
2       0.378 45 1.0      0.310     0.715    0.244     0.329    0.420  0.0448
3       0.356 45 1.0      0.310     0.715    0.240     0.323    0.411  0.0437
4       0.333 45 1.0      0.310     0.715    0.236     0.316    0.403  0.0429
5       0.311 45 1.0      0.310     0.715    0.230     0.310    0.396  0.0424
6       0.311 45 0.0      0.233     0.715    0.179     0.256    0.341  0.0415
7       0.289 45 0.0      0.233     0.715    0.175     0.249    0.331  0.0400
8       0.267 45 0.0      0.233     0.715    0.171     0.243    0.323  0.0388
9       0.244 45 0.0      0.233     0.715    0.166     0.237    0.315  0.0380
10      0.244 45 1.0      0.310     0.715    0.210     0.291    0.379  0.0432
11      0.222 45 0.0      0.233     0.715    0.161     0.230    0.308  0.0377
12      0.222 45 0.0      0.233     0.715    0.161     0.230    0.308  0.0377
13      0.222 45 0.0      0.233     0.715    0.161     0.230    0.308  0.0377
14      0.222 45 1.0      0.310     0.715    0.202     0.285    0.375  0.0441
15      0.222 45 1.0      0.310     0.715    0.202     0.285    0.375  0.0441
16      0.200 45 0.0      0.233     0.715    0.155     0.224    0.302  0.0377
17      0.178 45 0.0      0.233     0.715    0.148     0.218    0.297  0.0381
18      0.156 45 0.0      0.233     0.715    0.140     0.211    0.292  0.0389
Mean          45 0.4      0.267     0.715    0.191     0.267    0.351  0.0410
\end{CodeOutput}
\end{CodeChunk}

The shrinkage estimates are the same for all players because all players have the same 45 at-bats, i.e., the same amount of the observed information. 

\begin{CodeChunk}
\begin{CodeInput}
R> summary(b.output)
\end{CodeInput}
\begin{CodeOutput}
Main summary:

                            obs.mean  n    X1 prior.mean shrinkage low.intv 
Unit with min(obs.mean)        0.156 45 0.000      0.233     0.715    0.140     
Unit with median(obs.mean)1    0.244 45 0.000      0.233     0.715    0.166     
Unit with median(obs.mean)2    0.244 45 1.000      0.310     0.715    0.210     
Unit with max(obs.mean)        0.400 45 1.000      0.310     0.715    0.248     
Overall Mean                         45 0.444      0.267     0.715    0.191     


                            post.mean upp.intv post.sd
                                0.211    0.292  0.0389
                                0.237    0.315  0.0380
                                0.291    0.379  0.0432
                                0.335    0.429  0.0462
                                0.267    0.351  0.0410

Second-level Variance Component Estimation Summary:
alpha=log(A) for Gaussian or alpha=log(1/r) for Binomial and Poisson data:

post.mode.alpha post.sd.alpha post.mode.r
          -4.73         0.957         113


Regression Summary:

      estimate    se  z.val p.val
beta1   -1.194 0.131 -9.129 0.000
beta2    0.389 0.187  2.074 0.038
\end{CodeOutput}
\end{CodeChunk}

The regression coefficient for the outfielder indicator is significant, considering that $p$~value for $\hat{\beta}_2$ (\code{beta2}) is 0.038. It means that the  two estimates for the expected random effects for the outfielders and infielders are significantly different. Also, the positive sign of $\hat{\beta}_{2}$ indicates that the population batting average for  outfielders tends to be higher than that for infielders. The  estimated odds ratio is $\exp(0.389)=1.48$.

\begin{CodeChunk}
\begin{CodeInput}
R> plot(b.output)
\end{CodeInput}
\end{CodeChunk}
\begin{figure}[h!]
\begin{center}
\includegraphics[width = 2.7in]{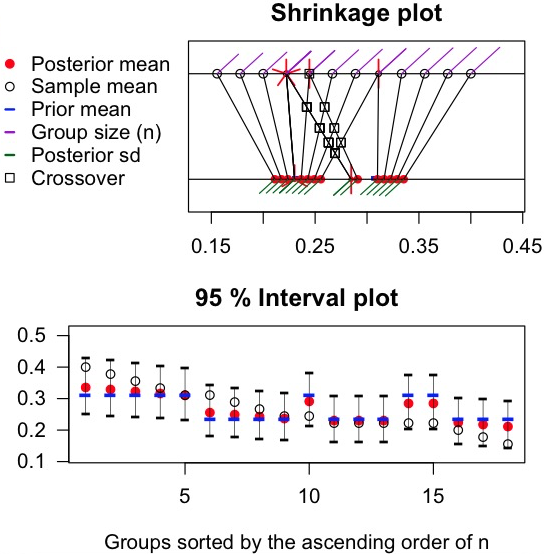}
\caption{Shrinkage plot and 95\% interval plot for 18 baseball players.}
\label{fig:baseball}
\end{center}
\end{figure}

The shrinkage plot in Figure \ref{fig:baseball} shows that  the observed batting averages (empty dots) on the upper horizontal line shrink towards the two expected random effects, 0.233 and 0.310. The short red line symbols near some empty dots are for when two or more points have the same mean and are plotted over each other. For example, five players (from the 11th player to the 15th) have the same batting average, 0.222, and at this point on the upper horizontal line, there are short red lines toward five directions.


The 95\% interval plot in Figure \ref{fig:baseball} shows the range of true batting average for each player, which clarifies the  regression towards the mean within two groups. The 10th, 14th, and 15th players, for example, are outfielders but their observed batting averages are far lower than the first five outfielders. This can  be attributed to their bad luck because their observed batting averages are close to the lower bounds of their interval estimates. The  regression towards the mean indicates that their batting averages  shrink towards the expected random effect of outfielders (0.310) in the long run.

To check the level of trust in these interval estimates, we  proceed to frequency method checking by assuming the estimates, 112.95 for $\hat{r}$ and (-1.194, ~0.389) for $\hat{\boldsymbol{\beta}}$, are the generative values. 

\begin{CodeChunk}
\begin{CodeInput}
R> b.coverage <- coverage(b.output, nsim = 1000) 
\end{CodeInput}
\end{CodeChunk}
\begin{figure}[h!]
\begin{center}
\includegraphics[width = 2.7in]{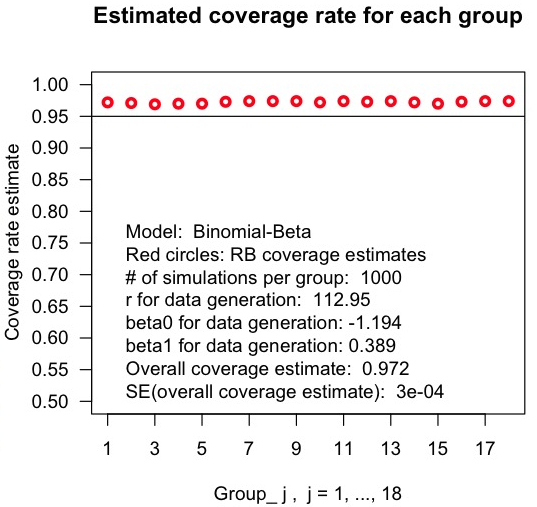}
\caption{Coverage plot via frequency method checking for 18 players.}
\label{fig:baseball2}
\end{center}
\end{figure}


In Figure \ref{fig:baseball2},  the estimated coverage probabilities for random effects are beyond 0.95, conservatively satisfying the 95\% confidence level if $r=\hat{r}$ and $\boldsymbol{\beta}=\hat{\boldsymbol{\beta}}$. The  overall unbiased coverage estimate across all the players is 0.972. 

We can check the RB unbiased coverage estimates and their standard errors for each player;
\begin{CodeChunk}
\begin{CodeInput}
R> bcv$coverageRB
\end{CodeInput}
%$
\begin{CodeOutput}
 [1] 0.971 0.973 0.972 0.972 0.970 0.973 0.973 0.974 0.973 0.973 0.971 0.973 
[13] 0.973 0.972 0.972 0.971 0.973 0.971
\end{CodeOutput}
\end{CodeChunk}
\begin{CodeChunk}
\begin{CodeInput}
R> bcv$se.coverageRB
\end{CodeInput}
\begin{CodeOutput}
 [1] 0.0015 0.0012 0.0013 0.0014 0.0016 0.0010 0.0012 0.0010 0.0010 0.0013 
[11] 0.0015 0.0013 0.0019 0.0013 0.0014 0.0015 0.0011 0.0014
\end{CodeOutput}
\end{CodeChunk}




If we want to draw 2,000 posterior samples of random effects and hyper-parameters from their full posterior distribution via the A-R method, we use the following \proglang{R} code.
\begin{CodeChunk}
\begin{CodeInput}
R> b.output <- gbp(y, n, x, model = "binomial", n.AR = 2000)
\end{CodeInput}
\end{CodeChunk}
The sampling result saved in \code{b.output} consists of 8,000 weights (\code{b.output$weight}), 2,000 posterior samples of $\alpha$ (\code{b.output$alpha}), a 2,000$\times2$ matrix of $\beta$ (\code{b.output$beta})  each column of which corresponds to 2,000 posterior samples of each regression coefficient, and a $k\times$2,000 matrix of random effects (\code{b.output$p}) each row of which has posterior samples of each random effect.

If we run the frequency method checking using this sampling result, \code{b.output}, obtained via the A-R method, the $N_{\textrm{sim}}$ simulations also run the A-R method each time.

\section[packageoverview]{Usage of functions in \pkg{Rgbp}}\label{sec5}
In this section, we describe more specific usage with various options of the two main functions of \pkg{Rgbp}, \emph{i.e.}, \code{gbp} for model fitting and \code{coverage} for frequency method checking. 

\subsection{Model fitting}\label{code_model}
The function \code{gbp} creates an S3 object ``gbp'' on which  three generic functions \code{plot}, \code{print}, and \code{summary} are defined. 

There are two cases according to whether covariates are available or not. When no covariates are available, the function \code{gbp} requires fitting an intercept term  or designating known values of   the expected random effects, \emph{i.e.}, the intercept term must be either estimated or known. The default of \code{gbp}  is to fit an intercept term. The value(s) of the known expected random effect(s) can be assigned through an optional argument \code{mean.PriorDist}. Note that \code{gbp} can fit the Poisson model only when the values of expected random effects, $\lambda^E_j$, are known. The usage of \code{gbp} to  fit each model without any covariates is 
\begin{CodeChunk}
\begin{CodeInput}
R> g.output <- gbp(y, se.or.n, model = "gaussian")
R> b.output <- gbp(y, se.or.n, model = "binomial")
R> p.output <- gbp(y, se.or.n, mean.PriorDist, model = "poisson")
\end{CodeInput}
\end{CodeChunk}

The argument \code{y} is a vector of $k$ observed sample means  for the Gaussian model, $k$ observed numbers of successful outcomes for the Binomial model, and $k$ observed outcome counts  for the Poisson model. The argument  \code{se.or.n} is a vector of $k$ standard errors of each sample mean for the Gaussian model, $k$ numbers of trials for the Binomial model, and $k$ exposures for the Poisson model. The argument \code{mean.PriorDist} is either a constant (if all the known expected random effects are the same) or a vector of $k$ known expected random effects.


If covariate information for each group is available, users can fit the Gaussian and Binomial models, using the following codes.
\begin{CodeChunk}
\begin{CodeInput}
R> g.output <- gbp(y, se.or.n, X, model = "gaussian")
R> b.output <- gbp(y, se.or.n, X, model = "binomial")
\end{CodeInput}
\end{CodeChunk}

The argument \code{X} is a matrix of covariate(s) each column of which corresponds to one covariate for $k$ groups. For example, if users have two covariates for each group, the argument \code{X} must be a $k\times2$ matrix to estimate three regression coefficients $\boldsymbol{\beta}=(\beta_1, \beta_2, \beta_3)$ including an intercept term, $\beta_1$, as a default. If users do not want to include an intercept term ($\beta_1=0$), estimating two regression coefficients for the two covariates, users can add an optional argument \code{intercept} as follows.
\begin{CodeChunk}
\begin{CodeInput}
R> g.output <- gbp(y, se.or.n, X, model = "gaussian", intercept = FALSE)
\end{CodeInput}
\end{CodeChunk}

The function \code{gbp} contains more optional arguments. The argument \code{confidece.lvl}, whose default value is 0.95, sets the confidence level, producing $100\times$\code{confidece.lvl}\% interval estimates for the random effects. For the Gaussian model, setting the argument \code{normal.CI} to \code{TRUE} lets \code{gbp}  use a Normal approximation to the unconditional posterior distribution of the random effect \citep{tang2011}. The default value of \code{normal.CI} is \code{FALSE} for the skewed-Normal approximation \citep{kelly2014advances}. 

The function \code{gbp} uses the A-R method to fit the Binomial model if users assign the desired number of posterior samples ($N$ in  \eqref{weight}) through the argument \code{n.AR}; its default value is zero. There are several arguments related to the A-R method. The argument \code{n.AR.factor} determines how many trial samples the method draws; its default value is four, meaning that the function \code{gbp} draws $4\times$\code{n.AR} trial samples and accepts or rejects them. The argument \code{trial.scale} is $\psi$ determining the scale parameter $\sigma$ of the skewed-$t$ distribution; its default value is 1.3. The argument \code{save.result} indicates whether \code{gbp} saves the whole posterior samples of the random effects and hyper-parameters; its default value is \code{TRUE}. The two arguments \code{t} and \code{u}, taking on non-negative and positive values, respectively, allow users to choose the joint hyper-prior density function, $f(r, \boldsymbol{\beta})\propto1/(t+r)^{u+1}$; the default values for \code{t} and \code{u} are 0 and 1, respectively, for the joint hyper-prior density function specified in  \eqref{eq:hyper}. 

For example, when there are two covariates, the following code produces 2,000 posterior samples of random effects and those of hyper-parameters, $r$ and $\boldsymbol{\beta}_{(3\times1)}$ including an intercept term, via the A-R method with 8,000 trial samples.
\begin{CodeChunk}
\begin{CodeInput}
R> b.output <- gbp(y, se.or.n, X, model = "binomial", n.AR = 2000)
\end{CodeInput}
\end{CodeChunk}

The object \code{b.output} above contains 8,000 weights (\code{b.output$weight}), 2,000 posterior samples of $\alpha$ (\code{b.output$alpha}), a 2,000$\times3$ matrix of $\beta$ (\code{b.output$beta})  each column of which corresponds to 2,000 posterior samples of each regression coefficient, and a $k\times$2,000 matrix of random effects (\code{b.output$p}) each row of which has posterior samples of each random effect.


The S3 object ``gbp'' benefits from three generic functions, \code{print},
\code{summary}, and \code{plot}. The estimation result for all the random
effects appears if users type the ``gbp'' object in the \proglang{R} console, which
plays the same role of the function \code{print} with its default argument
\code{sort = TRUE}. When the argument \code{sort} is set to \code{TRUE}, the
function \code{print} displays the estimation result for all the groups in the
ascending order of $n$ for the Binomial and Poisson model and the descending
order of standard errors for the Gaussian model. When the argument \code{sort}
is \code{FALSE}, the estimation result is returned in the order of data input.
\begin{CodeChunk}
\begin{CodeInput}
R> b.output
R> print(b.output, sort = FALSE)
\end{CodeInput}
\end{CodeChunk}

The function \code{summary} prints a detailed estimation result, including the estimation result for the hyper-parameters, $A$ (or $r$) and $\boldsymbol{\beta}$.
\begin{CodeChunk}
\begin{CodeInput}
R> summary(b.output)
\end{CodeInput}
\end{CodeChunk}

The function \code{plot} draws a shrinkage plot and $100\times$\code{confidece.lvl}\% interval plot for random effects, see Figure \ref{fig:hospshr}, \ref{fig:8schoolsplot}, or \ref{fig:baseball} for example. Its default argument ``\code{sort = TRUE}'' displays the $100\times$\code{confidece.lvl}\% interval plot in the ascending order of $n$ for the Binomial and Poisson model and the descending order of standard errors for the Gaussian model. When the argument \code{sort} is set to \code{FALSE}  the $100\times$\code{confidece.lvl}\% interval plot is displayed in the order of data input.
\begin{CodeChunk}
\begin{CodeInput}
R> plot(b.output)
R> plot(b.output, sort = FALSE)
\end{CodeInput}
\end{CodeChunk}

\subsection{Frequency method checking}
The function \code{coverage} conducts the frequency method checking. It estimates
the coverage properties for our estimators of the random effects at
a particular value of the hyperparameters by averaging the coverage over many simulated datasets. The basic usage of \code{coverage} needs a ``gbp'' object, such as \code{b.output} above, as the first argument;
\begin{CodeChunk}
\begin{CodeInput}
R> cov <- coverage(b.output, nsim = 1000)
\end{CodeInput}
\end{CodeChunk}

The argument \code{nsim} sets the number of simulations, $N_{\textrm{sim}}$, defined in Section \ref{data_generation}. If users do not assign values of the hyper-parameters through the arguments \code{A.or.r} and \code{reg.coef}, then the function \code{coverage} automatically sets the estimated posterior modes of hyper-parameters saved in the ``gbp'' object (or their posterior medians if the acceptance-rejection method for the Binomial model is used) as the generative values of hyper-parameters. If users want to conduct the frequency method checking with different generated values of hyper-parameters, for example, $r=100$ and $\boldsymbol{\beta}=(2, 5)^\top$ when one covariate was used with an intercept term, then users can specify them via the arguments \code{A.or.r} and \code{reg.coef};
\begin{CodeChunk}
\begin{CodeInput}
R> cov <- coverage(b.output, A.or.r = 100, reg.coef = c(2, 5), nsim = 1000)
\end{CodeInput}
\end{CodeChunk}

When users fit a model via \code{gbp} with known expected random effects, e.g., a Poisson model with known values of $\{\lambda_1, \lambda_2, \ldots, \lambda_k\}$, \code{coverage} conducts the frequency method checking based on these known values as a default. However, users may want to conduct the frequency method checking with different known values of the expected random effects. For example, if users want to try  a different value of the expected random effect, \emph{e.g.}, $\lambda_j=30$ (or can be a vector of different values), the argument \code{mean.PriorDist} is added as follows.
\begin{CodeChunk}
\begin{CodeInput}
R> cov <- coverage(p.output, mean.PriorDist = 30, nsim = 1000)
\end{CodeInput}
\end{CodeChunk}
The resulting frequency method checking is based on the estimated posterior mode of $r$ (because it is not specified through \code{A.or.r}) and the newly specified value of the expected random effect, 30, as its known value.

Though the function \code{coverage} does not produce an S3 object,  the result of \code{coverage} contains various numerical details; $k$ RB coverage estimates (\code{cov$coverageRB}) and their standard errors (\code{cov$se.coverageRB}),    overall unbiased coverage estimate (\code{cov$overall.coverageRB}) and its standard error (\code{cov$se.overall.coverageRB}), etc.

A coverage plot summarizing the result of \code{coverage} automatically appears. If the result is saved in a variable such as \code{cov} above, then users can  recall the coverage plot, using the function \code{coverage.plot}.
\begin{CodeChunk}
\begin{CodeInput}
R> coverage.plot(cov)
\end{CodeInput}
\end{CodeChunk}

\section[Discussion]{Discussion} \label{discussion}
\pkg{Rgbp} is an \proglang{R} package for estimating and validating two-level Gaussian, Poisson, and Binomial hierarchical models. The package aims to provide a procedure that is computationally efficient with good frequency properties and includes frequency method checking functionality to examine repeated sampling properties and to test that the method is valid at specified hyper-parameter values.

As an alternative to other maximization based estimation methods such as MLE and REML, \pkg{Rgbp} provides approximate point and interval estimates of parameters via ADM. Using the ADM approach, with our specified choice of priors, protects from cases of overshrinkage and undercoverage from which the aforementioned methods suffer  \citep{accuracy1988}.

A benefit of \pkg{Rgbp} is that it produces non-random output (except the A-R method for the Binomial model) and so results are easily reproduced and compared across studies. In addition to being a stand-alone analysis tool the package can be used as an aid in a broader estimation procedure. For example, by checking the similarity of output of \pkg{Rgbp} and that of another estimation procedure such as MCMC (Markov chain Monte Carlo), the package can be used as a confirmatory tool to check whether the alternative procedure has been programmed correctly. In addition, the parameter estimates obtained via \pkg{Rgbp} can be used to initialize an MCMC thus decreasing time to convergence. Lastly, due to its speed and ease of use, \pkg{Rgbp} can be used as a method of preliminary data analysis. Such results may tell statisticians and practitioners alike whether a more intensive method in terms of implementation and computational time, such as MCMC, is needed.


\appendix
\section{Posterior propriety of the Poisson model}\label{propriety_poisson}
If the posterior distribution of $r$ is proper, then the full posterior distribution of random effects and $r$ is also proper because 
\begin{equation}
f(\boldsymbol{\lambda}, r\mid \boldsymbol{y})=f(\boldsymbol{\lambda}\mid \boldsymbol{y})\times f(r\mid \boldsymbol{y}),
\end{equation} 
where $f(\boldsymbol{\lambda}\mid \boldsymbol{y})$ is a product of $k$ proper conditional posterior density function in \eqref{gammapost}. Thus, our goal is to show that $\int_0^\infty f(r\mid \boldsymbol{y})dr<\infty$:
\begin{align}
f(r\mid\boldsymbol{y})&\propto \frac{1}{r^2}L(r)\propto\frac{1}{r^2}\prod^{k}_{j=1} \frac{\Gamma(r \lambda^E_j+y_j)}{\Gamma(r\lambda^E_j)}(1-B_{j})^{y_{i}}B_{j}^{r \lambda^E_j}\\
&=\frac{1}{r^2}\left[r^{\sum_{j=1}^ky_j}+\cdots+a_k r^{k}\right]\exp\left(-r\sum_{j=1}^k\lambda^E_j\log(1+n_j/r)\right)\prod_{j=1}^k\left(\frac{n_j}{n_j+r}\right)^{y_j} \label{final_post_proper},
\end{align}
where the polynomial function of $r$ in the bracket has constant coefficients. 

If there are at least two groups whose observed values $y_j$ are non-zero, then $f(r\mid\boldsymbol{y})$ goes to zero as $r$ approaches zero due to the polynomial function of $r$ in \eqref{final_post_proper}; the following two factors in \eqref{final_post_proper} approach one. As $r$ becomes infinite, $f(r\mid\boldsymbol{y})$ touches zero exponentially fast due to the exponential term in the middle of \eqref{final_post_proper}. Thus, the integration of $f(r\mid\boldsymbol{y})$ must be finite.

\section*{Acknowledgments}
The authors thank the editor and reviewers for their insightful comments that substantially improved this work, Professor Cindy Christiansen, Professor Phil Everson and the 2012 class of Harvard's Stat 324r: Parametric Statistical Inference and Modeling for their valuable inputs, and Steven Finch for his careful proofreading.

\bibliography{jss1274}

\end{document}